\documentclass{emulateapj}

\usepackage{psfig}
\usepackage{amsmath}
\usepackage{amssymb}
\usepackage{graphicx}
\usepackage{appendix}

\newcommand{\lSect}[1]{{\label{sec:#1}}}
\newcommand{\lFig}[1]{{\label{fig:#1}}}
\newcommand{\lEq}[1]{{\label{eq:#1}}}
\newcommand{\lTab}[1]{{\label{tab:#1}}}
\newcommand{\Msun}{\ensuremath{\mathrm{M}_\odot}}

\newcommand{\Zsun}{\ensuremath{\mathrm{Z}_\odot}}

\newcommand{\Teff}{\ensuremath{T_{\mathrm{eff}}}}
\newcommand{\Jtot}{\ensuremath{J_{\mathrm{tot}}}}
\newcommand{\MBH}{\ensuremath{M_{\mathrm{BH}}}}
\newcommand{\Menv}{\ensuremath{M_{\mathrm{env}}}}
\newcommand{\Mdisk}{\ensuremath{M_{\mathrm{disk}}}}
\newcommand{\Mgrav}{\ensuremath{M_{\mathrm{grav}}}}
\newcommand{\Mpresn}{\ensuremath{M_{\mathrm{pre-SN}}}}
\newcommand{\Mhecore}{\ensuremath{M_{\mathrm{He-core}}}}
\newcommand{\tauc}{\ensuremath{\tau_{\mathrm{circ}}}}
\newcommand{\taus}{\ensuremath{\tau_{\mathrm{sync}}}}
\newcommand{\lambdas}{\ensuremath{\lambda_{\mathrm{sync}}}}
\newcommand{\jcrit}{\ensuremath{j_{\mathrm{crit}}}}
\newcommand{\vesc}{\ensuremath{v_{\mathrm{esc}}}}
\newcommand{\aHe}{\ensuremath{a_{\mathrm{He}}}}
\newcommand{\BE}{\ensuremath{B\!E}}
\newcommand{\rhob}{\ensuremath{\bar{\rho}}}

\newcommand{\kms}{\ensuremath{\mathrm{km}\,\mathrm{s}^{-1}}}
\newcommand{\ergs}{\ensuremath{\mathrm{erg}\,\mathrm{s}^{-1}}}
\newcommand{\rads}{\ensuremath{\mathrm{rad}\,\mathrm{s}^{-1}}}
\newcommand{\junit}{\ensuremath{\mathrm{cm}^2\,\mathrm{s}^{-1}}}
\newcommand{\Junit}{\ensuremath{\mathrm{erg}\,\mathrm{s}}}
\newcommand{\Iunit}{\ensuremath{\mathrm{g}\,\mathrm{cm}^{-2}}}
\newcommand{\erg}{\ensuremath{\mathrm{erg}}}
\newcommand{\cm}{\ensuremath{\mathrm{cm}}}
\newcommand{\ms}{\ensuremath{\mathrm{ms}}}
\newcommand{\Sec}{\ensuremath{\mathrm{s}}}
\newcommand{\K}{\ensuremath{\mathrm{K}}}
\newcommand{\yr}{\ensuremath{\mathrm{yr}}}
\newcommand{\AU}{\ensuremath{\mathrm{AU}}}
\newcommand{\Msuns}{\ensuremath{\Msun\,\mathrm{s}^{-1}}}
\newcommand{\gccm}{\ensuremath{\mathrm{g}\,\mathrm{cm}^{-3}}}

\newcommand{\D}{\mathrm{d}}
\newcommand{\DyDx}[2]{\frac{\D #1}{\D #2}}
\newcommand{\DDx}[1]{\frac{\D}{\D #1}}
\newcommand{\Mdot}{\dot{M}}
\newcommand{\tauff}{\ensuremath{\tau_{\mathrm{ff}}}}
\newcommand{\B}[1]{\left(#1\right)}
\newcommand{\Bx}[2]{\B{#1}^{\!#2}}
\newcommand{\BB}[2]{\B{\frac{#1}{#2}}}
\newcommand{\BBx}[3]{\Bx{\frac{#1}{#2}}{#3}}

\newcommand{\FIGFF}[2]{{\ref{fig:#2}{#1}}}

\newcommand{\FIG}[2]{{Fig.~\FIGFF{#1}{#2}}}
\newcommand{\Fig}[1]{{\FIG{}{#1}}}

\newcommand{\Sectff}[1]{{\ref{sec:#1}}}
\newcommand{\Sect}[1]{{\S~\Sectff{#1}}}

\newcommand{\Appendix}[1]{{Appendix~\Sectff{#1}}}
\newcommand{\Eqref}[1]{{\ref{eq:#1}}}
\newcommand{\Eqff}[1]{{(\Eqref{#1})}}

\newcommand{\EQ}[1]{{Equation~\Eqff{#1}}}

\newcommand{\Tab}[1]{{Table~\ref{tab:#1}}}

\begin{document}

\title{Long Gamma-Ray Transients from Collapsars}

\author{S. E. Woosley\altaffilmark{1} and Alexander Heger\altaffilmark{2}}

\altaffiltext{1}{Department of Astronomy and Astrophysics, University
  of California, Santa Cruz, CA 95064; woosley@ucolick.org}

\altaffiltext{2}{Minnesota Institute of Astrophysics, 
School of Physics \& Astronomy, University of Minnesota, Twin Cities,
Minneapolis, MN 55455; alex@physics.umn.edu}

\begin{abstract}
In the collapsar model for common gamma-ray bursts, the formation of a
centrifugally supported disk occurs during the first $\sim$10 seconds
following the collapse of the iron core in a massive star. This only
occurs in a small fraction of massive stellar deaths, however, and
requires unusual conditions. A much more frequent occurrence could be
the death of a star that makes a black hole and a weak or absent
outgoing shock, but in a progenitor that only has enough angular
momentum in its outermost layers to make a disk. We consider several
cases where this is likely to occur - blue supergiants with low mass
loss rates, tidally-interacting binaries involving either helium stars
or giant stars, and the collapse to a black hole of very massive
pair-instability supernovae. These events have in common the accretion
of a solar mass or so of material through a disk over a period much
longer than the duration of a common gamma-ray burst. A broad range of
powers is possible, $10^{47}$ to $10^{50}\,$erg s$^{-1}$, and this
brightness could be enhanced by beaming. Such events were probably
more frequent in the early universe where mass loss rates were
lower. Indeed this could be one of the most common forms of gamma-ray
transients in the universe and could be used to study first generation
stars. Several events could be active in the sky at any one time. A
recent example of this sort of event may have been the SWIFT transient
Sw-1644+57.
\end{abstract}


\section{INTRODUCTION}
\lSect{intro}

The collapsar model \citep{Woo93} was originally proposed as an
explanation for common, ``long-soft'' gamma-ray bursts (GRBs) with
durations near 20 s. This is a typical time scale for the
gravitational collapse of the carbon-oxygen core of a massive star and
the congruence of these time scales was an argument in favor of the
model.  It is well known, however, that forming a disk so quickly
requires an unusually large amount of angular momentum to be stored in
the core up to the point when the star dies
\citep[e.g.,][]{Woo06}. This is presumably one reason why GRBs are
rare compared with supernovae.

Black holes are expected to form in a significant fraction of massive
star deaths \citep[e.g.,][]{Oco11} and it seems possible that, in at
least some of these, the black hole will form without generating an
outgoing shock. What observable signal would accompany such an event?
Would they all be ``un-novae'' \citep{Koc08}, stars that simply
suddenly disappear from view? Or will some sort of transient display
announce the birth of any black hole? Here we explore the possibility
that these black hole systems form an accretion disk, but only after
almost all of the star has accreted into the hole. Torques in
differentially rotating contracting stars tend to concentrate angular
momentum in the star's outer layers. The surface will also have a
large angular momentum in a tidally locked system. On the other hand,
stellar mass loss preferentially removes these same layers, so, just
as in the usual collapsar model, reducing mass loss, e.g., by reducing
the metallicity, favors the retention of high angular momentum at the
stellar surface.

Because the outer layers of the star take longer to fall in - 
months to years in the case of a red supergiant - the duration of the transient
produced by accretion through a disk in these cases is much longer
than for the common GRBs studied so far. They might even be confused
with other forms of gamma-ray transients like those arising from
super-massive black hole accretion in active galaxies
\citep{Woo11,Bur11,Lev11,Qua11}.

Here we shall consider four varieties of long duration transients
categorized by the kind of system where they occur: 1) single
supergiants with metallicity less than about $0.1\,\Zsun$ and low mass
loss (\Sect{BSG}); 2) red or blue supergiants in tidally-locked, or
nearly tidally-locked binaries (\Sect{RSG}); 3) pair-instability
induced collapse to a black hole (\Sect{pair}); and 4) helium stars in
tidally locked binaries (\Sect{helium}). In each case, the duration of
the transient that is produced is correlated with the radius of the
presupernova star, hence shortest for the helium stars, while the
luminosity is inversely correlated with the radius, hence faintest for
the red supergiants. As with common GRBs, we are unable to predict
\textsl{ab initio} the spectrum of these transients or the efficiency for
converting black hole accretion rate into electromagnetic
radiation. Presumably, as in other forms of transients powered by very
rapidly accreting, rotating black holes, accretion energy and black
hole spin energy power relativistic bi-polar outflows. Poynting flux,
shocks within these outflows, and shocks with the circumstellar medium
make x-rays and gamma-rays. It is common to take an efficiency of $1\,\%$
to $10\,\%$ times $\Mdot c^2$ for the efficiency \citep[e.g.,][]{McK06}
thus suggesting an integrated power of $\sim10^{52}\,\erg - 10^{53}\,\erg$
per solar mass accreted.  Interestingly these ``failed supernovae''
\citep{Woo93} are far more brilliant and powerful than their
(model-wise) more successful cousins, ordinary supernovae.

The idea that late time accretion in supernovae that make black holes
might power long gamma-ray (or hard x-ray) transients was suggested by
\citet{Mac01}. The transients discussed there, referred to as ``Type
2 collapsars'', were produced because of fall-back in an explosion
that launched a relatively strong outgoing shock in a red supergiant
progenitor. Two dimensional models explored jet propagation and break
out for a variety of fall-back rates, but little attention was given
to the angular momentum distribution necessary to form a disk. Here we
consider a wider variety of progenitors and pay closer attention to
the angular momentum, especially in the outer layers. \citet{Li03}
also discussed the possibility of very-long duration X-ray transients
powered by fall back and accretion through a disk. He suggested these
might look like ``ultra-luminous x-ray sources'' seen in nearby
galaxies. These would be much longer in duration (thousands of years)
and have much lower luminosity than the events considered here, but
are an example of the same general idea.

More recently \citet{Qua11} and \citet{Woo11} suggested that fallback
or a failed explosion in a red supergiant could give a long duration
gamma-ray transient that might be confused with a gamma-ray
blazar. Quataert and Kasen, in particular, suggest this as an
alternate interpretation of SWIFT transient Sw-1644+57 \citep{Bur11}.
Woosley's paper was accepted prior to the discovery of this event.

A variety of other models may lead to long duration bursts. For
example a white dwarf \citep{Fry99a} or helium star merging with a
black hole or neutron star \citep{Fry98,Zha01}.  A neutron star or
black hole merging with a a giant star might give a Thorne-Zytkow
object \citep{Ter95,Pod07} or a black hole accreting at a rapid rate
\citep{Che93}. Helium star merger with a compact object also gives a
long transient of some sort \citep{Fry05}. Here we consider only those
transients coming from stars that are not actively merging with a
companion.

\section{The Production of Stellar Mass Black Holes and Type 3 Collapsars}
\lSect{black}

A black hole will form inside a massive star when a sufficient
concentration of matter and energy accumulates to form an event
horizon. The simplest possibility is that some portion of a massive
star experiences an instability and falls freely, or almost freely,
until it reaches such a high density that a trapped surface is
formed. This rarely happens in nature though. Only in the most massive
stars experiencing a pair instability \citep{Heg02} or a general
relativistic instability \citep{Ful86,Mon11} can a trapped surface
form before the center reaches nuclear density and bounces. That is,
black hole formation in a massive star is almost always preceded by
protoneutron star formation \citep{Oco11}.

For stars in the range $\sim20\,\Msun - 100\,\Msun$, however, the
protoneutron star that is formed experiences rapid accretion from the
surrounding layers of heavy elements - silicon, oxygen, etc. If
rotation, magnetic fields and neutrinos do not readily produce an
outgoing shock, the hot protoneutron star grows beyond a critical mass
and collapses to a black hole. The accretion rate is larger for more
massive helium cores \citep{Fry99}, and black hole formation without
an outgoing shock may be a frequent occurrence in stars with main
sequence masses over about $20\,\Msun$ \citep{Oco11,Sma09}. Whether
{\sl all} stars above about $20\,\Msun$ produce black holes without
outgoing shock waves is uncertain, but the possibility is not
precluded by observations or theory.  ``Prompt'' black hole formation
is probably more likely in very low metallicity stars that have
experienced little mass loss and thus have especially large helium
cores, but it could occur for solar metallicity stars, depending upon
the uncertain mass loss rate \citep{Smi11}.

It is also possible that such a weak outgoing shock is generated that
most of the star fails to explode and a black hole is made by fall
back \citep{Zha08}. These were the systems explored by \citet{Mac01}.
The ones treated here differ in assuming that any outgoing shock is
insufficient to eject more than a small fraction of the surface layers
of the star. We shall call such events, where only the surface layers
have enough rotation to form a disk and a shock so weak that these
layers are not ejected ``Type 3 Collapsars''. This is to distinguish
them from ordinary GRBs where the disk forms promptly from material
deep inside a highly differentially rotating helium core \citep[Type
  1;][]{Woo93,Mac99} and from systems where longer duration transients
result from the delayed fall back of other material also deep inside
the star \citep[Type 2;][]{Mac01,Heg03}.

\section{Examples of Type 3 Collapsars}
\lSect{stars}

The preservation of angular momentum in the surface layers of the star
can occur either because mass loss is inefficient in a single star or
because of tidal interaction with a close companion. Here we consider
examples of both sorts.

\subsection{Massive Stars with Low Metallicity and Low Mass Loss}
\lSect{BSG}

A paucity of metals in a star affects its evolution in several ways -
reduced energy generation by the CNO cycle in the hydrogen burning
shell, reduced opacity, and reduced mass loss. Of these effects, the
reduction in mass loss is least well understood, but the lack of
atomic lines in the atmosphere on the main sequence and of grains as a
red giant seems certain to lead to some reduction. A common assumption
is that the mass loss for massive main sequence stars depends upon
some fractional power of the metallicity \citep{Kud02}.  More
important and less certain is the dependence of mass lost as a red or
blue supergiant on metallicity.  It is thought that mass loss from
cool giants is more dependent upon pulsations and grain formation
\citep{Rei77,Smi11}. One might expect therefore a rapid fall off in
mass loss below some value necessary for significant grain
production. It has been estimated that red giant mass loss will be
significantly less below $0.1\,\Zsun$ \citep{Bow91,Zij04}.

If stars do not lose mass then they conserve angular momentum. Since
the natural course of evolution leads to the contraction and spin up
of the inner star, shear instabilities and magnetic torques will
concentrate an increasing amount of angular momentum in the outer
regions of the star.  Whereas the inner part does spin faster due to
contraction, it actually loses most of its initial angular momentum by
the time it reaches central carbon burning. \citet{Heg10} surveyed the
evolution of non-rotating zero-metal massive stars from $10\,\Msun$
to $100\,\Msun$. In a parallel study currently underway
\citep{Heg11}, we are surveying the evolution of rotating stars in the
same mass range with metallicity $0$, $10^{-3}$ and $10^{-1}$ that of
the sun. The two stars discussed here are extracted from that survey
and are typical in the way that they accumulate large amounts of
angular momentum in their outer layers before dying as
supergiants. All calculations of stellar evolution presented here used
the Kepler code \citep{Wea78,Woo02,Heg00} and included angular
momentum transport by magnetic torques \citep{Spr02}. Models V24 and
V36 are $24\,\Msun$ and $36\,\Msun$ main sequence stars with metallicity
$0.1\,\%$ solar. Both stars rotated rigidly on the main sequence with a
moderate speed of about $200\,\kms$ which is about $20\,\%$ of the
Keplerian value.  Both presupernova stars had hydrogenic envelopes
that were, throughout most of their mass, radiative. Model V36,
however, had a low mass surface convection zone that included $0.11
\,\Msun$. Model V36 was consequently a yellow supergiant at death ($L =
2.2 \times 10^{39}\,\ergs$; $R = 5.4 \times 10^{13}\,\cm$;
$\Teff = 5700\,\K$). Model V24 was a rather large blue supergiant
($L = 1.1 \times 10^{39}\,\ergs$, $R = 1.0 \times 10^{13}\,\cm$;
$\Teff = 11,300\,\K$).

These two stars ended their lives with cores of helium and heavy
elements of $8.3\,\Msun$ and $15.0\,\Msun$ respectively and angular
momentum distributions as shown in \Fig{bsg}. Whereas the lack of
sufficient angular momentum within the helium core precludes making a
disk around a black hole within that mass, there is ample rotation in
the outer part of the hydrogen envelope to do so.  This could be a
very common occurrence. From the survey of \citet{Heg11}, the total
angular momentum of a massive star at birth that has an equatorial
rotation speed of about $20\,\%$ Keplerian on the main sequence is
\begin{equation}
\Jtot = I \omega \approx 2 \times 10^{52} \left(\frac{M}{\Msun}\right)^{\!1.8}\,\Junit.
\end{equation}
If the star does not lose mass, most of this angular momentum becomes
concentrated, in the presupernova star, in a nearly rigidly rotating
hydrogenic envelope with radius either $\lesssim 10^{13}\,\cm$ (blue
supergiant) or $\sim10^{14}\,\cm$ (red supergiant). Whereas the
density declines with radius in the actual envelope, one can obtain
some interesting scaling relations by assuming constant density. For a
moment of inertia, $I \approx 0.4 M R^2$, the angular velocity at the
surface, $R = 10^{14}\,R_{14}\,\cm$, is
\begin{equation}
\omega \approx 1.2 \times 10^{-10} \, R_{14}^{-2} \left(\frac{
  M}{20\,\Msun}\right)^{\!0.8} \rads.
\end{equation}
The specific angular momentum of this rigidly rotating envelope is $j
= \frac{2}{3} \omega r^2$, or about $10^{18}\,\junit$ at the surface
of a red supergiant and $10^{20}\,\junit$ for a blue supergiant, if no
mass is lost.

These can be compared with the angular momentum required to make a
disk around a non-rotating black hole, $j = 2 \sqrt{3} GM/c = 3.1
\times 10^{17} (\MBH/20\,\Msun)\,\junit$ or $j = 2/\sqrt{3} GM/c =
1.0 \times 10^{17} (\MBH/20\,\Msun)\,\junit$ for a maximally rotating
hole (Kerr parameter $a=1$). All massive stars that do not lose mass
have sufficient angular momentum in their outermost layers to make a
disk around any black hole formed in their collapse. 

Of course stars do lose mass and the fact that, for constant $\omega$,
$j \propto r^2$ means that the first mass to be lost contains most of
the angular momentum. In practice, we find that a red supergiant that
loses more than a few percent of its total mass before dying will
probably not make a black hole-disk system.  

This sensitivity to mass loss is illustrated by the evolution of a $25
\,\Msun$ model, O25, with $10\,\%$ solar metallicity in a calculation
that included mass loss at one-half the standard value for that
metallicity. The star lost $1.17\,\Msun$ ($5\,\%$) of its mass and
died as a red supergiant with a radius of $1.1\times10^{14}\,\cm$. The
value of $j$ in its outermost layers was $2.8\times10^{17}\,\junit$,
too little to form a disk around a Schwarzschild black hole, but
possibly just enough to form a disk around a mildly rotating one.  The
point where $j$ declined to one-half its surface value in the
presupernova star was located $5.5\,\Msun$ interior. This is
borderline case.  Stars with less mass loss, i.e., lower metallicity,
and stars that never become red giants could make long gamma-ray
transients (\Tab{models}).

\subsection{Supergiants in Interacting Binaries}
\lSect{RSG}

Based upon theoretical models for single massive stars, the red
supergiant progenitors of common Type IIp supernovae are not expected
to have rapidly rotating envelopes. Expansion to a red giant slows the
surface layers. Torques transfer the angular momentum of the more
rapidly rotating inner regions outwards concentrating it in the
outermost layers. For stars of sufficient metallicity, stellar winds
are efficient at removing these layers and, along with them, most of
the star's angular momentum. For example, a $25\,\Msun$ solar
metallicity star rotating rigidly with an equatorial speed of
$200\,\kms$ on the main sequence ends up a $12.1\,\Msun$ supernova
progenitor if commonly used estimates for mass loss are employed
\citep{Woo07}. The surface rotation rate of that red supergiant (R =
$9 \times 10^{13}\,\cm$) is $2 \times 10^{-12}\,\rads$ and its angular
momentum at the surface is $1.6 \times 10^{16}\,\junit$. Nowhere in
the star does any matter have enough angular momentum to form a disk
around a black hole.  If a black hole forms without making an outgoing
shock, the star abruptly disappears. This sort of ``un-nova'' might be
a common occurrence in the modern universe \citep{Koc08}, though see
\citet{Smi11}.

The outcome can be quite different in a binary system, however. A
stellar merger could lead to a rapidly rotating envelope
\citep{Iva03}, but such mergers probably lead to ejection of a
common envelope. An interesting alternative is detached binary systems
with a small enough separation for tidal interaction, but too distant
to merge or exchange mass.

The circularization and synchronization time scales for binary stars,
one of which is a red giant, have been considered by \citet{Zah66},
\citet{Zah89} and \citet{Ver95}. For a primary with mass, $M$,
envelope mass $\Menv$, luminosity $L$, and radius $R$, in an orbit
having radius $d$, with a companion of mass $M_2$, the {\sl
  circularization} time, $\tauc$ is \citep{Ver95,Phi92}
\begin{equation}
\frac{1}{\tauc} = f \left(\frac{L}{\Menv R^2}\right)^{\!1/3} \,
\frac{\Menv}{M} \frac{M_2}{M} \frac{M+M_2}{M} \left(\frac{R}{d}\right)^{\!8}\!,
\lEq{circ}
\end{equation}
where $f$ is a number close to unity. The lifetime of a $25\,\Msun$
main sequence star as a red supergiant is approximately
$700,000\,\yr$, depending upon uncertain convection parameters.  The
mass of the envelope that is convective at any one time, however, is
only about $3\,\Msun$ and the star loses more than this in its
lifetime. A more appropriate time scale might be the time the star
spends losing its last $3\,\Msun$, which is about $250,000\,\yr$. The
mass of the star when it becomes a supernova is $12.1\,\Msun$ and we
assume a companion mass of, e.g., $6\,\Msun$. The luminosity of the
star as a red supergiant is $1.0 \times 10^{39}\,\ergs$ and its moment
of inertia is approximately $4 \times 10^{61}\,\Iunit$. \EQ{circ} then
implies that during its last $250,000\,\yr$, orbits with radius $d
\lesssim 3.8 R \approx 18\,\AU$ will circularize.

The \emph{synchronization} timescale for co-rotation is somewhat
different \citep{Zah77,Zah89} and, in the present case, is easier to
achieve. 
\begin{equation}
\frac{1}{\taus} = 6 q^2 \lambdas \left(\frac{L}{\Menv
  R^2}\right)^{\!1/3} \, \frac{MR^2}{I} \, \left(\frac{R}{d}\right)^{\!6},
\end{equation}
Most of the angular momentum is in the hydrogen envelope and, from the
same $25\,\Msun$  model, $MR^2/I = 2.9$. With $q \approx 1$ and
$\lambdas\approx 0.02$ \citep{Zah89},
\begin{equation}
\frac{1}{\taus} \approx 0.1 \left(\frac{L}{\Menv R^2}\right)^{\!1/3}
\, \left(\frac{R}{d}\right)^{\!6},
\end{equation}
which only requires $d\lesssim 6.7 R$ for synchronization in
$250,000\,\yr$.  In such a system, the period will be shorter than
about $30$ years ($\omega = 10^{-8}\,\rads$).  Systems much closer will
merge or experience mass exchange, so we take this to be a typical
rotation period for the envelope in a detached binary system that
might achieve corotation. Longer period systems will still transfer
significant rotation to the envelope, however.

This is about four orders of magnitude more angular momentum in the
outer envelope than in the case of the single star. Since the outer
several solar masses of the presupernova star is convective, this
rotation rate will persist to some depth. All of that mass would have
sufficient angular momentum to form an accretion disk if its helium
core collapsed to a black hole. Even then though, the total angular
momentum of the envelope, $2.0 \times 10^{53}\,\Junit$, would be a
small fraction of the orbital angular momentum of the binary pair of
stars. The addition of only $1\,\%$ of the angular momentum estimated
here would still result in large quantities of material forming a
disk, so the range of binary separations allowed is actually larger,
perhaps by a factor $\sim(100)^{1/6}$, or about two.

Since the time scale for material to fall in from $10^{13} -
10^{14}\,\cm$ is longer than any realistic viscous time scale for the
disk, the accretion rate will be given by the collapse timescale of
the envelope. For matter at the base of the hydrogen shell where a
disk would first form, the enclosed mass is $9.3\,\Msun$ and the
radius about $1 \times 10^{13}\,\cm$. At the outer edge of the star
the mass is $12.1\,\Msun$ and the radius, $9.8 \times
10^{13}\,\cm$. The corresponding free fall times ($\tauff = \left(24
  \pi G \bar \rho\right)^{-1/2} =
446\,\left(\bar{\rho}/1\,\gccm\right)^{-1/2}\,\Sec$) are $2.1 \times
10^5\,\Sec$ and $5.7 \times 10^6\,\Sec$ suggesting that the event
would last for months.  Accreting $3\,\Msun$ in $10^6\,\Sec$ with
$10\,\%$ efficiency for converting accreted mass to outgoing energy
would give a jet power of $10^{48}\,\ergs$. Conversion of only
$10\,\%$ of the jet power ($1\,\%$ of the accreted mass energy) into
gamma-rays would explain the recently discovered transient Sw 1644+57
\citep{Bur11}, especially if moderate beaming were involved.

Whereas the focus here is on red supergiants because they give the
longest transients, tidal locking would also occur in blue supergiants
with smaller radii.  The key point is that forced co-rotation with a
body in Keplerian orbit gives, for a large range of orbital
separations, envelope material that is rotating too fast to accrete,
without hindrance, into a stellar mass black hole. Since the time
scale for accretion is set by the collapse time for the outer part of
the star, not the viscous time of the disk, a transient from a tidally
locked blue supergiant would be similar in duration within a factor of
a few to those studied for single blue supergiants in \Sect{BSG}.
Blue supergiants, however, will have less mass in the outer layers of
the envelope and hence should have less powerful outbursts
(\Appendix{append} and \citealt{Qua11}).

\subsection{Pair-Instability in Very High Mass Stars}
\lSect{pair}

Following helium burning, non-rotating helium cores with masses
greater than $133\,\Msun$ will collapse directly into black holes with
no outgoing shock \citep{Heg02}. The collapse is caused by the
pair-instability and, above this mass, nuclear burning is unable to
reverse the implosion. Rotation, unless it is very rapid and highly
differential, does not change this mass limit greatly since the
instability exists in the deep interior where the ratio of centrifugal
force to gravity is small.

The collapse of a rotating pair unstable star to a black hole was
followed in two dimensions by \citet{Fry01}. First, a $300\,\Msun$
main sequence star was evolved to the point of instability in the
KEPLER (1D) code. This star produced a helium core of $180\,\Msun$. The
initial star was assumed to rotate rigidly on the main sequence with a
ratio of equatorial speed to Keplerian of 20\%. Angular momentum
transport, especially by shear mixing and Eddington Sweet circulation,
was followed throughout the evolution, but magnetic torques were
neglected. As a result, the collapsing helium core rotated
sufficiently rapidly to produce a disk outside of a massive black hole
once about $140\,\Msun$ of the $180\,\Msun$ core of helium and heavy
elements had collapsed. \citet{Fry01} speculated that the accretion of
the remaining $30\,\Msun - 40\,\Msun$ might produce a very bright,
lengthy GRB.

We have recalculated the evolution of similar stars, but with magnetic
torques \citep{Spr02} included in the simulation. Models Z250A, B, and
C are stars with zero metallicity and a mass at the time they formed
of $250\,\Msun$.  They differ in the amount of angular momentum each
had on the main sequence. Z250A had $0.75 \times 10^{54}\,\Junit$
corresponding to a rotation rate of $170\,\kms$, or $12\,\%$ Keplerian.
Models Z250B and Z250C had $1.0 \times 10^{54}\,\Junit$ and $1.5
\times 10^{54}\,\Junit$ of angular momentum and rotational speeds of
$220\,\kms$ ($15\,\%$ Keplerian) and $310\,\kms$ ($19\,\%$
Keplerian). These made helium cores of $143\,\Msun$, $166\,\Msun$, and
$222\,\Msun$ respectively, all of which collapsed directly to black
holes following helium depletion.  Unlike in \citet{Fry01} however,
none of these helium cores had sufficient angular momentum to make a
disk around a black hole anywhere in their interior. Because mass loss
was neglected in these models, the angular momentum lost from the
helium core due to magnetic coupling instead ended up concentrated in
the outer part of the hydrogen envelope.

\Fig{pair} shows the evolution of the angular momentum in Model Z250B.
The evolution of the other two models was similar. Inadequate angular
momentum exists in the helium core to form a disk but, for Model
Z250B, the outer $16\,\Msun$ can still do so. As in the tidally locked
red supergiant case, the collapse time scale for this material is days
to months, so a high energy transient of around $10^{47}\,\ergs$,
possibly boosted by beaming, could exist for that time.

It should be noted, however, that mass loss is even more uncertain in
these models than in the others.  Due to rotationally induced mixing,
the envelopes of all three models were appreciably enhanced in C, N,
and O and thus were red supergiants when they died.  Neglecting mass
loss in such a situation is probably not justified. Also there may be
other mechanisms for losing mass in such massive stars besides line
and grain driven mass loss \citep{Smi06}.

\subsection{Helium stars in Binaries}
\lSect{helium}

Helium stars in close tidally locked binaries have been frequently
suggested as progenitors for common GRBs
\citep{Tut03,Pod04,Izz04,Bog07,Van07}. On the positive side, such
events should occur frequently in nature and, if the core of a massive
Wolf-Rayet star collapses to a black hole while the surface is forced
to rotate at a fraction of its Keplerian speed, disk formation around
a black hole is assured. On the negative side, however, the production
of a GRB with typical duration $\sim$10 s, requires that the
\emph{inner} part of the helium star rotate rapidly, not just its
surface. The inclusion of magnetic torques in tidally locked stars
leads, by itself, to too little rotation to make a disk in that part
of the star which might accrete rapidly \citep{Woo06,Van07}. On the
other hand, leaving out magnetic torques makes it difficult to produce
the spin rates of ordinary pulsars \citep{Heg05}, and could lead to an
\emph{overabundance} of GRB progenitors.

This is not to say that close binary systems do not produce
\emph{some} form of gamma-ray transient though. The angular momentum
in their cores may be adequate to produce a millisecond magnetar, and
thus a common GRB, just not a (Type 1) collapsar. If a black hole
forms, the accretion of the \emph{outer} layers of the star may lead
to the formation of a disk and GRB much longer than typical. It is
this latter possibility that we explore here.

Following \citet{Van07}, we consider the evolution of helium cores of
$8\,\Msun$ and $16\,\Msun$ whose surface s have become tidally locked
with a closely orbiting companion.  \citet{Van07} found the shortest
possible periods were 2.05 hours ($\omega = 8.5 \times
10^{-4}\,\rads$) and $2.47$ hours ($\omega = 7.1 \times
10^{-4}\,\rads$) for $8\,\Msun$ and $16\,\Msun$ respectively, and they
gave examples of systems containing a compact object and a helium core
that might approach these limiting values.  An $8\,\Msun$ He core in a
tidally locked binary with a $0.8\,\Msun$ companion filling its Roche
lobe had a longer period of $7.17$ hours ($\omega = 2.4 \times
10^{-4}\,\rads$). Here we calculate two models for each helium core
mass. Model 8A has an enforced rotation rate in its outer $0.5\,\Msun$
of $2 \times 10^{-4}\,\rads$ throughout its evolution.  Models
8B, 16A, and 16B have rotation rates of $8.5 \times 10^{-4}\,\rads$, $2
\times 10^{-4}\,\rads$, and $7 \times 10^{-4}\,\rads$ respectively.

\Fig{binaryw} shows the evolution of the angular momentum distribution
as a function of mass for these four models. During helium burning,
magnetic coupling maintains a state of nearly rigid rotation with an
angular speed given by the surface boundary condition. For both Model
8A and 16A, this corresponds to a ratio of rotational speed to
Keplerian of $\sim8\,\%$. For Models 8B and 16B, the ratio is
$\sim32\,\%$.  Roughly the inner $75\,\%$ ($8\,\Msun$) or $85\,\%$
($16\,\Msun$) of the helium star is convective during helium burning and
rigid rotation is enforced in that region. The outer part of the star
is radiative, but magnetic torques there suppress differential
rotation. During carbon burning and more advanced stages, the inner
core contracts and rotates more rapidly. Magnetic torques are too
inefficient to enforce rigid rotation. Later, during carbon burning,
the stars ignite a convective helium burning shell which persists
until the star dies. This convective shell does not extend all the way
to the surface though, but does extend beyond the $0.5\,\Msun$ at the
surface where corotation is enforced. Thus all stars die with an
angular velocity in their helium convective shell given by the surface
boundary condition, but have larger rotation rates in their cores, due
to contraction during advanced burning stages.

\Fig{binaryj} shows the corresponding angular momentum in the evolving
models. Shear instabilities and magnetic torques transport angular
momentum out of the inner core and to the surface. In the absence of a
binary companion, the surface would have actually spun up, but would
have been removed by mass loss. In the present parametrized
calculations though, there is no mass loss and the set rotation rate
of the surface absorbs any excess angular momentum delivered to
it. Consequently the total angular momentum of the star decreases with
time. For Model 8A at helium ignition, it was $9.0 \times
10^{51}\,\Junit$, but for the presupernova star it was $1.6 \times
10^{50}\,\Junit$.

\Fig{binaryjfin} shows the final distribution of angular momentum in
the stars compared with that required to make a disk around a black
hole, once the mass interior to the given sample point has collapsed.
Four curves are given. Three are for the minimum angular momentum
required to form a disk at the last stable orbit of a) a non-rotating
(Schwarzschild) black hole; b) a Kerr black hole; and c) a black hole
with rotation given by the actual angular momentum distribution of the
presupernova star. The fourth curve is the angular momentum
distribution in the presupernova star. Because the inner parts of the
star are not rotating very rapidly, curve a) and c) are similar except
near the surface. Only where the actual angular momentum lies above
curve c) can a disk form. Material with less angular momentum plunges
directly into the hole.

None of the models has enough angular momentum to form a disk in its
inner core. Thus a Type~1 collapsar will not occur. The angular
momentum in the inner $1.7\,\Msun$ of the stars is still appreciable,
however. Models 8A and 16A would form $1.4\,\Msun$ (gravitational
mass) pulsars with periods of $7.4\,\ms$ and $2.6\,\ms$
respectively. Models 8B and 16B would form pulsars of $3.2\,\ms$ and
$1.5\,\ms$.  Model 16B thus qualifies as a ``millisecond magnetar''
candidate and might power a common GRB if a pulsar is able to form
before accretion turns it into a black hole. Models 8B, 16A and 16B
could also make powerful pulsar-powered supernovae.  Model 8A and
other more slowly rotating systems probably would not.

Of greater interest for the present paper, however, is the large
angular momentum in the \emph{outer} layers of three of the
models. These would be relevant if the center of the star collapsed to
a black hole rather than a millisecond pulsar. Model 8A has angular
momentum sufficient to form a disk in its outer $0.074\,\Msun$. Two
other models (\Tab{models}) have similar large rotation in their
surface layers. This is consistent with the results of \citet{Van07}
who found that their most rapidly rotating $8\,\Msun$ and $16\,\Msun$
helium cores ended their lives with $j$ in outer layers of $3.7 \times
10^{17}\,\junit$ and $6.0 \times 10^{17}\,\junit$. The radius and
gravitational potential of the high angular momentum surface layers
allows an order of magnitude estimate of the free fall time scale
(\Tab{models}), about $100\,\Sec$.

The tidal interaction actually acts to brake the rotation of the
surface layers compared to what they would have had in a star without
mass loss and no companion star. To illustrate this, Model 8C was
calculated, starting from the rigidly rotating helium burning stage of
Model 8B (total angular momentum $3.6 \times 10^{51}\,\Junit$), but
with no surface boundary condition and no mass loss. That is the
rotation rate of the surface layers was allowed to adjust to be
consistent with whatever angular momentum was transported to
them. Without mass loss, the star was forced to conserve angular
momentum overall and the very outer layers ended up rotating very
rapidly (\Fig{binaryjfin}).  The angular speed, rather than being
$\omega = 7.5 \times 10^{-4}\,\rads$ was $3.7 \times
10^{-3}\,\rads$. In fact the outer tenth of a solar mass rotated so
rapidly that it would be centrifugally ejected. In a realistic
calculation, this matter and more underlying matter would probably
have been ejected as a disk or a centrifugally boosted wind. The deep
interior still lacked sufficient angular momentum to form a disk
around a black hole, but the amount of surface material that could
form a disk (if mass loss did not remove it) was significant, about 2
\Msun, comparable to that in Model 8B.  Two other models like 8C were
calculated that included mass loss appropriate for a Wolf-Rayet star
at solar metallicity and one-tenth that value. The first ended up with
a final mass of $3.95\,\Msun$ and no surface layers with sufficient
angular momentum to form a disk. The second ended up with a mass of
$7.09\,\Msun$ and still had enough angular momentum in its outer solar
mass to make a disk.

So the major role of the close companion in the calculations done here
is to preserve the angular momentum in the outer layers that might
otherwise be removed by mass loss. A companion may also be essential
so that the helium core rotates rapidly in the first place, following
the loss of the star's hydrogen envelope. When magnetic torques are
included in the calculation, the helium cores of supernova progenitors
that are red and blue supergiants rotate too slowly for any portion of
that core to make a disk. A close companion could either preserve
angular momentum in the core by removing the envelope very early in
the evolution or add angular momentum by tidal locking later on.

\section{Conclusions}
\lSect{conclude}

\Tab{models} and \Tab{models1} summarize the wide variety of
transients that can be produced by Type~3 collapsars. In \Tab{models}
$\Mdisk$ is the amount of mass capable of forming a centrifugally
supported disk around a black hole with a Kerr parameter given by
accreting all the mass interior to $\Mdisk$ and $R$ is the range of
stellar radii where that mass exists. The accretion time scale has
been estimated using the radius for the center of mass of $\Mdisk$.
The luminosity has been estimated by multiplying the effective
accretion rate (mass/time scale) by $1\,\%\, c^2$. This assumes
$10\,\%$ efficiency for making a jet and $10\,\%$ efficiency for
converting jet energy into hard radiation. Most models produce extreme
Kerr black holes ($a=1$, \Tab{models1}), so the radiative
efficiencies may be higher than assumed here \citep{Tch11}.  The
numbers in the table are only order of magnitude estimates.

\Tab{models1} gives the the estimated pulsar period formed in the
stellar collapse assuming the angular momentum that exists in the
inner $1.7\,\Msun$ of the core, a final gravitational mass of
$1.4\,\Msun$, and a moment of inertia of $0.35\Mgrav R^2$
\citep{Heg05}. If a black hole forms before the protoneutron star has
radiated away most of its gravitational binding energy, the rotation
rate will be slower as will the rotational kinetic energy available
for explosion. $\Mpresn$ and $\Mhecore$ are the masses of the
presupernova star and its helium core respectively and $M(j > \jcrit)$
is the Lagrangian mass coordinates of the stellar layers that can form
a centrifugally supported disk. The Kerr parameter, $a$, is what would
develop if the entire presupernova, including its rapidly rotating
outer layers, collapsed to a black hole. In case where it exceeds
unity it has been capped at $1$.  The Kerr parameter for the helium
core alone is $\aHe$.  $\BE(j > \jcrit)$ is the net binding energy, in
units of $10^{49}\,\erg$, of the material that could form a disk.

A range of masses can form disks and a large range of time scales
characterize the accretion, hence the possible luminosities span many
orders of magnitude.  The shorter brighter transients from Wolf-Rayet
stars in tidally locked binaries might comprise an interesting subset
of long, but otherwise ordinary GRBs, as others have noted. This
channel could be especially important in regions where high
metallicity and high mass loss preclude the production of GRBs by
solitary stars.  The mass of $^{56}$Ni produced in the collapsar model
is sensitive to the temperature, density, time scale, and ejection
fraction of the disk and is difficult to estimate. The lower accretion
rate, larger black hole mass, and smaller Kerr parameter for the
models presented here, however, will give lower temperatures and
densities in the disk \citep{Pop99}.  Below about $0.01\,\Msuns$, the
nature of the accretion changes and the disk is no longer neutrino
dominated. A thicker disk might qualitatively change the nature of the
transient (MacFadyen, private communication). It is thus reasonable to
expect that the $^{56}$Ni production, per solar mass of material
accreted, will be different here and perhaps much smaller.  These
bursts might not be accompanied by supernovae \citep[e.g.,][]{Gal06}.

On the longer end, blue and red supergiants will produce transients of
order $10^4\,\Sec$ and $10^5\,\Sec$ respectively. These could result
from either tidally locked systems or more distant systems with low
metallicity and mass loss. These fainter and potentially more common
systems may not have been detected yet and certainly should be
searched for. They could easily be confused with flaring gamma-ray
blazars \citep{Woo11,Bur11,Qua11,Lev11}, but would differ in that they
only happen once and would be associated with regions of massive star
formation, not particularly galactic nuclei. They too would probably
not be accompanied by supernovae and might have a different beaming
factor and spectrum from ordinary GRBs.  The black holes produced in
these supergiant progenitors would frequently be rotating at near the
maximum value (\Fig{bsg} and \Fig{rsg}).

An exciting prospect is that the first generations of stars, those
with metallicities from zero to $0.1$ solar might be studied directly
using the hard x-rays they emit when they die. Whereas predictions are
still clouded by the vagaries of mass loss and the explosion
mechanism, our studies predict that the fraction of massive stars
emitting such bursts could be substantial, essentially limited only by
the fraction that make black holes promptly. If none make black holes
promptly, then the death of every massive star must make a supernova,
though perhaps a faint one. This too is an interesting possibility
worth further exploration.

Because of their long duration and potentially high event rate, it is
possible that several Type~3 collapsars will be visible in distant
galaxies at any one time.  \citet{Mad98} estimate $20$ core-collapse
supernovae occur each year in a field of view $4\arcmin\times4\arcmin$ in
the redshift range $z = 0 \ldots 4$. This is about $6$ supernovae per
second spread over the whole sky. The observation of GRBs beyond
redshift $8$ shows that massive stars were dying at an even earlier
time. Black hole formation probably occurs in a significant fraction
of these supernovae \citep{Oco11,Sma09}, perhaps all those with mass
over $20\,\Msun$ and metallicity less than $10\,\%$ solar. In order to
estimate, crudely, an event rate, assume that prompt black hole
formation occurs in $10\,\%$ of core collapses and $1\,\%$ of these
stars either have such low mass loss rates or close companions that
their outer layers make a disk. Then $0.1\,\%$ of all massive star
deaths would make Type~3 collapsars.  For low metallicity stars, the
latter could be an underestimate. The universal supernova rate of $6$
per second then implies a gamma-ray transient event rate of $0.006$
per second. If each event lasted $10^4\,\Sec$, there would $\sim50$
events in progress at all times.  Most of these would be at high
redshift (low metallicity) and the spectrum would be softer and
duration longer than in the lab frame, so the number of active sources
could be higher.  On the other hand, beaming could increase the
brightness and decrease the event rate significantly.

In an extreme limiting case where $10\,\%$ of all massive star deaths
made Type~3 collapsars with duration $10^5\,\Sec$ with negligible
beaming, one would have $\sim 10^5$ active sources at any one time, or
about $1$ per square degree. If each source radiated $10^{51}\,\erg$,
this could potentially contribute to a ``diffuse'' cosmological
background of soft gamma-rays.

The greatest uncertainty in these results is whether the matter with
high angular momentum will get ejected, by mass loss either before the
explosion or during the explosion, without falling to the center of
the star. As \Tab{models1} shows, the binding energy of the critical
material is very much less than the kinetic energy of a typical
supernova ($\sim10^{51}\,\erg$), so a weak explosion or even strong
pulsations during silicon burning could eject it. It seems unlikely
that this would happen all of the time in all sorts of models, so
perhaps this enters in chiefly as a major uncertainty in the event
rates. We also note that the converse problem, the optical appearance
of a ``supernova'' exploding with only $\sim10^{49}\,\erg -
10^{50}\,\erg$ of kinetic energy is an interesting one too. One way or
another, it seems unlikely that these stars will all disappear without
a trace.

\acknowledgements

The authors appreciate valuable discussions with Christian Ott
regarding core-collapse in massive stars and with Ron Taam on tidal
interaction in close binaries.  This research has been supported by at
UCSC by the National Science Foundation (AST 0909129) and the NASA
Theory Program (NNX09AK36G). AH acknowledges support from the DOE
Program for Scientific Discovery through Advanced Computing (SciDAC;
DE-FC02-09ER41618), by the US Department of Energy under grant
DE-FG02-87ER40328, by NSF grant AST-1109394, and by the Joint
Institute for Nuclear Astrophysics (JINA; NSF grant PHY02-16783).

\newpage

\begin{appendix}

\section{Estimate of Accretion Rate}
\lSect{append}

Starting from the usual definition of free-fall time,
$$\tauff = 1/\sqrt{24 \pi G \rhob}$$
we obtain an order-of-magnitude estimate of the accretion rate from
$$
\Mdot 
=\Bx{\DyDx{\tauff}{m}}{-1} 
=\sqrt{18G}\Bx{\DDx{m}\BBx{m}{r^3}{-1/2}}{-1}
=\sqrt{18G}\Bx{-\frac12\BBx{m}{r^3}{-3/2}\B{\frac1{r^3}-\frac{3m}{r^4}\DyDx{r}{m}}}{-1}
$$
$$
\Mdot 
= 6\BBx{2Gm^3}{r^3}{1/2}\Bx{\frac{3m}{4\pi r^3\rho}-1}{-1}
= 6\BBx{2Gm^3}{r^3}{1/2}\Bx{\frac{\rhob}{\rho}-1}{-1} 
= 6\,m\BBx{8\pi G\rhob}{3}{1/2}\BB{\rho}{\rhob-\rho} 
$$
\begin{equation}
\Mdot
= \frac{2 m}{\tauff}\BB{\rho}{\rhob-\rho} 
\end{equation}
where we used $\rhob\!\B{m} = 3m/\B{4\pi r^3}$ and $\D m = 4\pi r^2
\rho\, \D r$, and $m$ is the mass coordinate and $r$ is the radius
coordinate of a shell.  Note that usually $\rhob\!\B{m}>\rho\!\B{m}$
in most parts of the star; density inversions can exist at the outer
edge of convective regions in mixing-length theory especially in RSGs
due to recombination and opacity drops.  If $\rho\ll\rhob$, as in the
case of the outer layers of a BSGs, we obtain
$\Mdot\approx2m\rho/(\tauff\rhob)\ll M/\tauff$, so BSGs may have a
more tenuous accretion rate and fainter, softer GRBs than RSGs based
on the estimate from $\tauff$.  On the other hand the average density
of the star is higher as it is more compact, which may be a slight
compensation.  Density inversions at the edge of RSG envelopes could
lead to spikes on accretion rate in this simple approximation.

\end{appendix}

\newpage


\clearpage

\begin{figure}
\begin{center}
\includegraphics[width=0.475\textwidth]{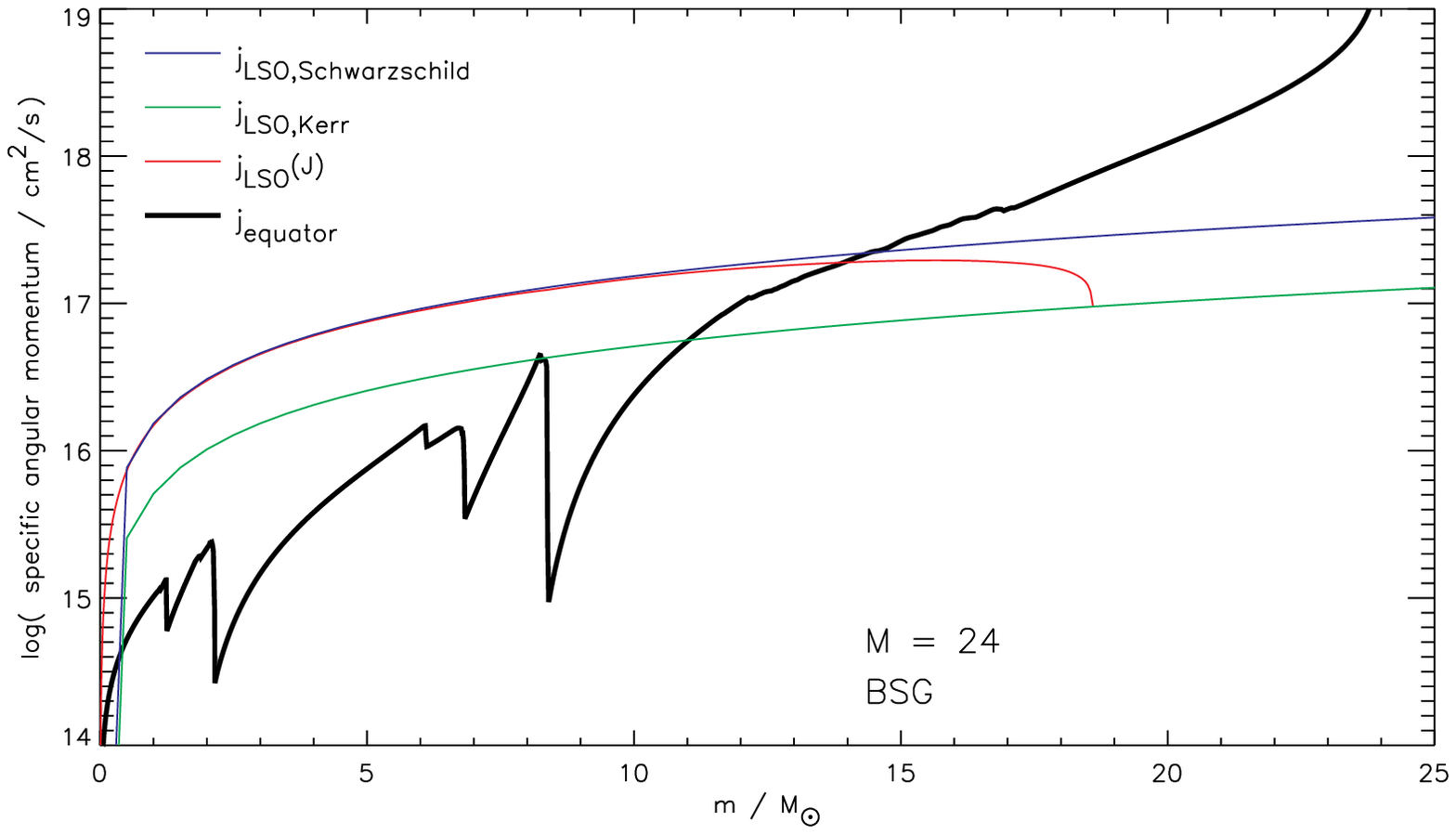}
\hfill
\includegraphics[width=0.475\textwidth]{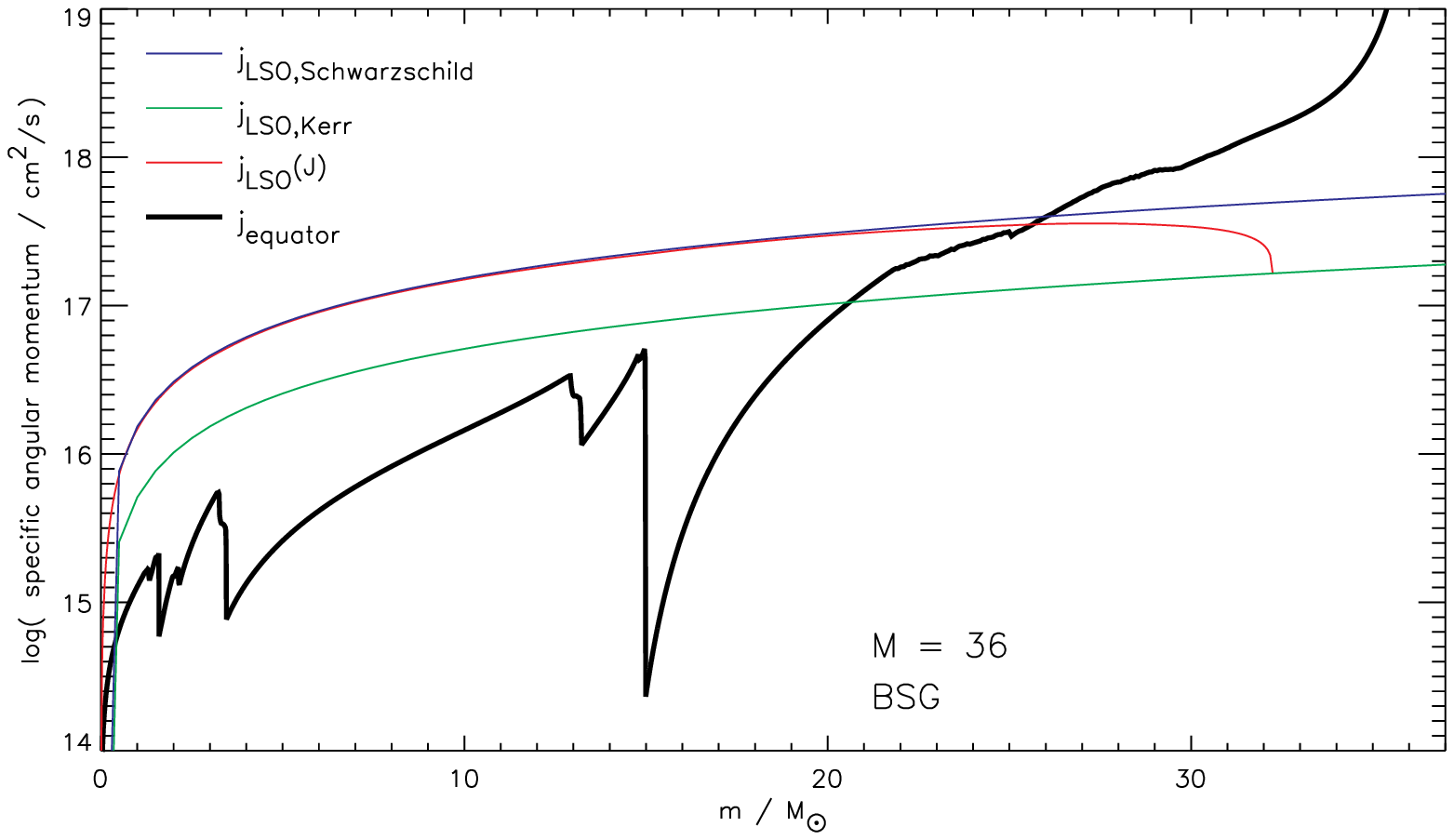}
\caption{Distribution of angular momentum with respect to mass in the
  presupernova models for V24 and V36. The smooth curves show the
  angular momentum required to form a stable disk at the last stable
  orbit of a Schwarzschild black hole (lower curve) or Kerr black hole
  (upper curve) including the given mass. The intermediate curve that
  follows the Schwarzschild curve until far out in the star uses the
  integrated angular momentum in the model to determine the last
  stable orbit. Where the irregular dark line showing the actual
  angular momentum on the star intersects this line a disk can
  form. The outer $9\,\Msun$ of Model V24 and the outer $10\,\Msun$ of
  Model V36 will form a disk. The edge of the helium cores of the two
  models is apparent in the sharp inflection in the angular momentum
  at $8.3\,\Msun$ and $15.0\,\Msun$. Mass loss was included in the
  calculation, but due to the low metallicity, only $0.05\,\Msun$ and
  $0.15\,\Msun$ was lost in V24 and V36, respectively. Note that, if
  all the surface material accreted here, the black hole would rotate
  at nearly its maximum allowed value (i.e., the red line intersects
  the green one at the end). \lFig{bsg}}
\end{center}
\end{figure}

\clearpage

\begin{figure}
\includegraphics[width=0.90\textwidth]{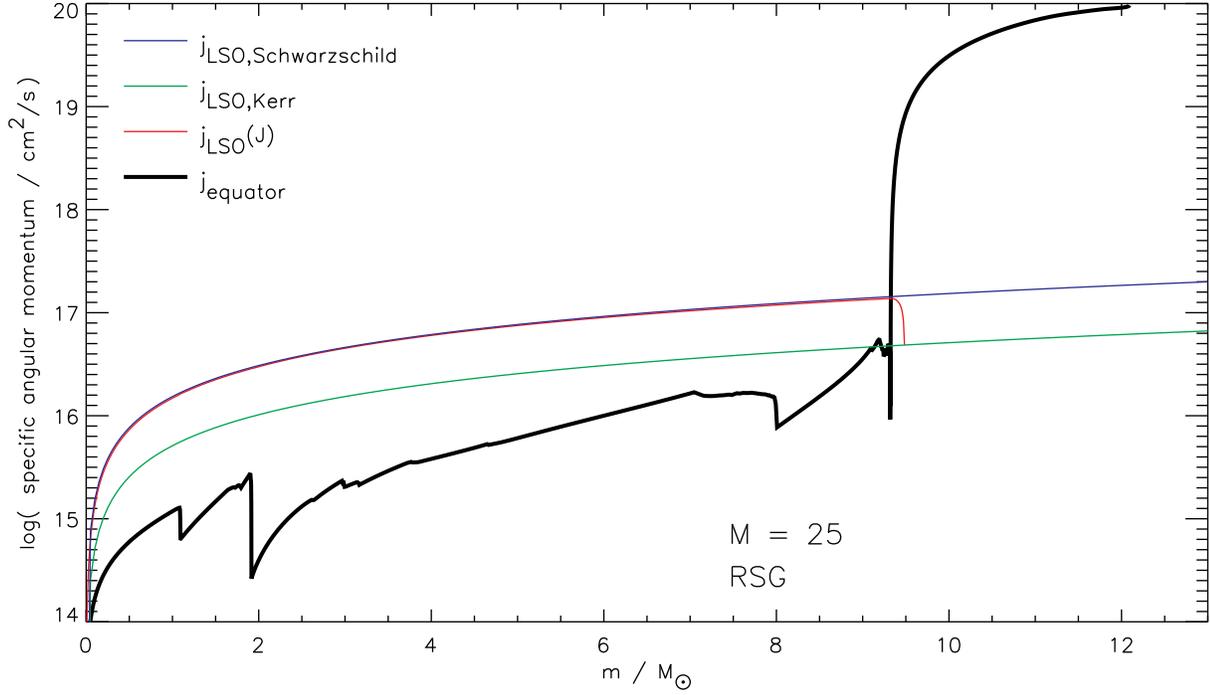}
\caption{Distribution of specific angular momentum in a tidally locked
  red supergiant in a binary system that has induced corotation of the
  convective envelope. As in \Fig{bsg}, the solid dark line shows the
  actual distribution of angular momentum while the lighter lines show
  the values required to form a disk around a non-rotating black hole,
  a Kerr black hole, or a black hole with the integrated angular
  momentum interior to the given mass. The angular velocity in the
  envelope is constant with $\omega = 10^{-8}\,\rads$. Note that the black
  hole formed here would be an extreme Kerr hole with $a \approx
  1$. \lFig{rsg}}
\end{figure}

\clearpage

\begin{figure*} 
\centering
\includegraphics[width=0.475\textwidth]{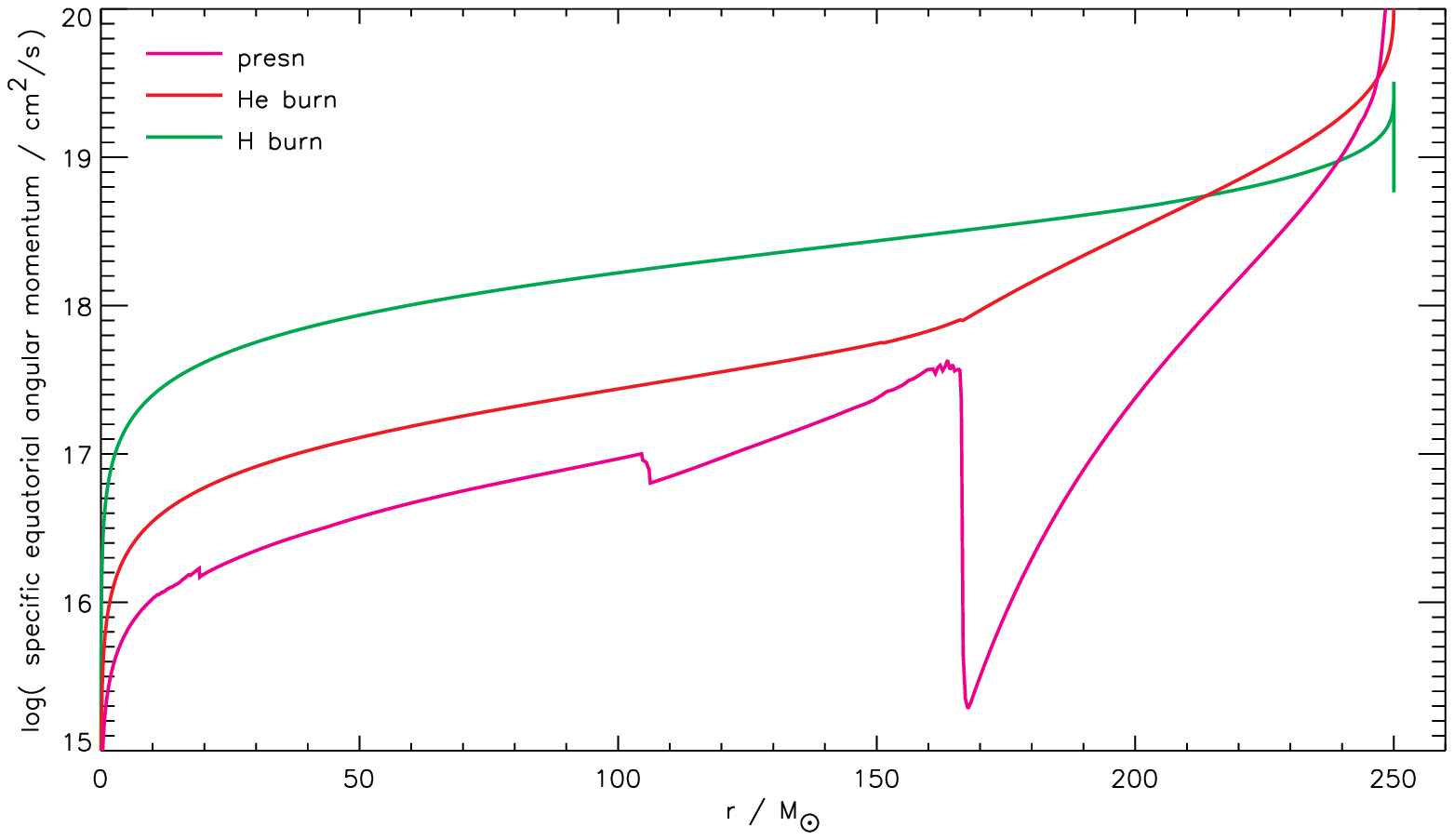}
\hfill
\includegraphics[width=0.475\textwidth]{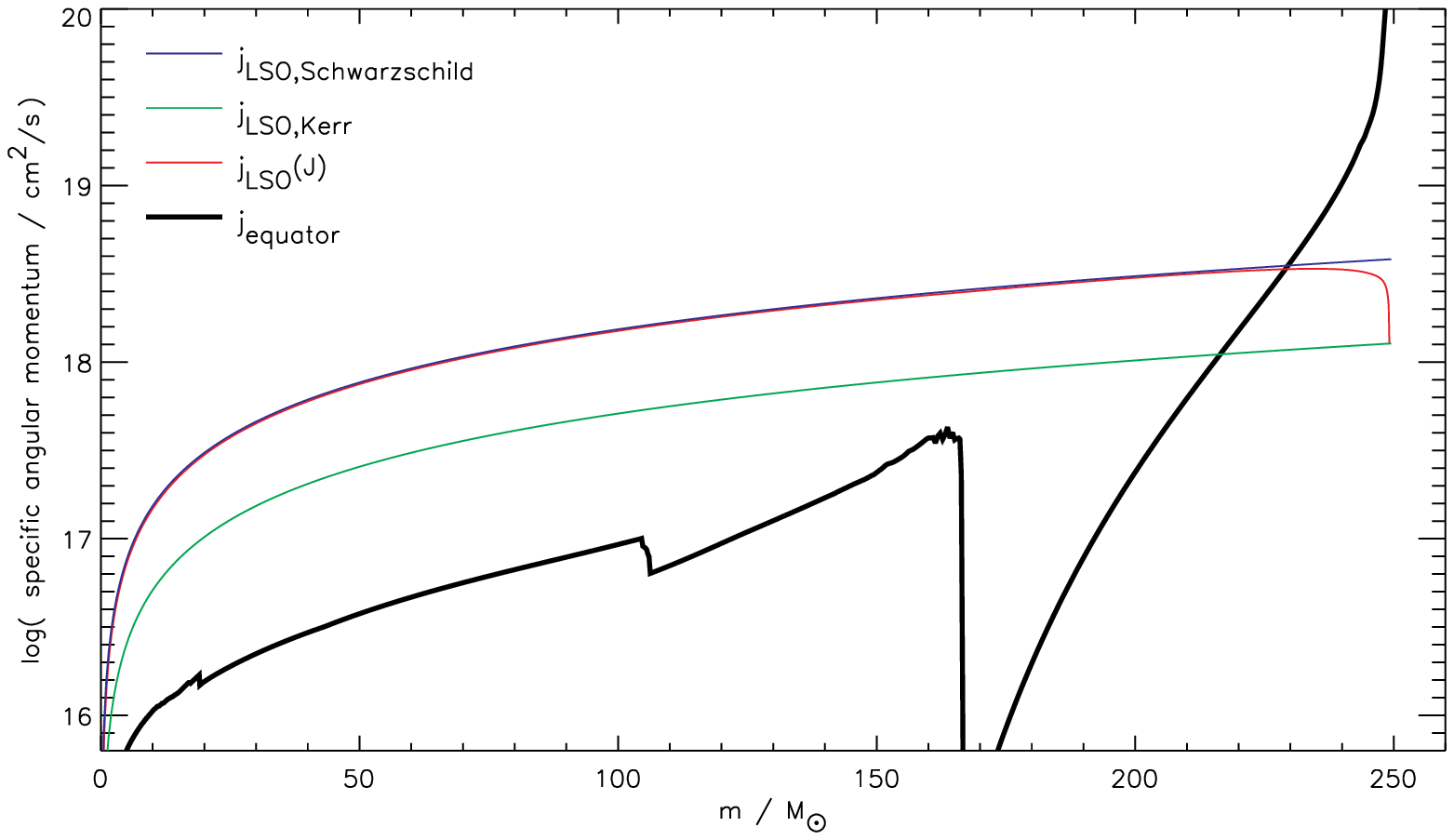}
\vskip 24pt
\caption{Final angular momentum for a $250\,\Msun$ pair-instability
  supernova (Model Z250B) in which rotation and magnetic torques were
  included in the calculation. The left panel shows the evolution of
  the angular momentum and the right panel shows the distribution of
  angular momentum in the presupernova star compared to what is need
  to make a disk around a black hole. The helium core mass is
  $166\,\Msun$ and only matter in the outer $16\,\Msun$ has sufficient
  angular momentum to make a disk. If the outer $16\,\Msun$ is not
  lost during the evolution a disk might still form and the black hole
  could end up rotating rapidly.  The radius of this matter is from
  $1.5 \times 10^{12}\,\cm$ to $5.9 \times 10^{13}\,\cm$.  \lFig{pair}}
\end{figure*}

\clearpage

\begin{figure*} 
\centering
\includegraphics[width=0.475\textwidth]{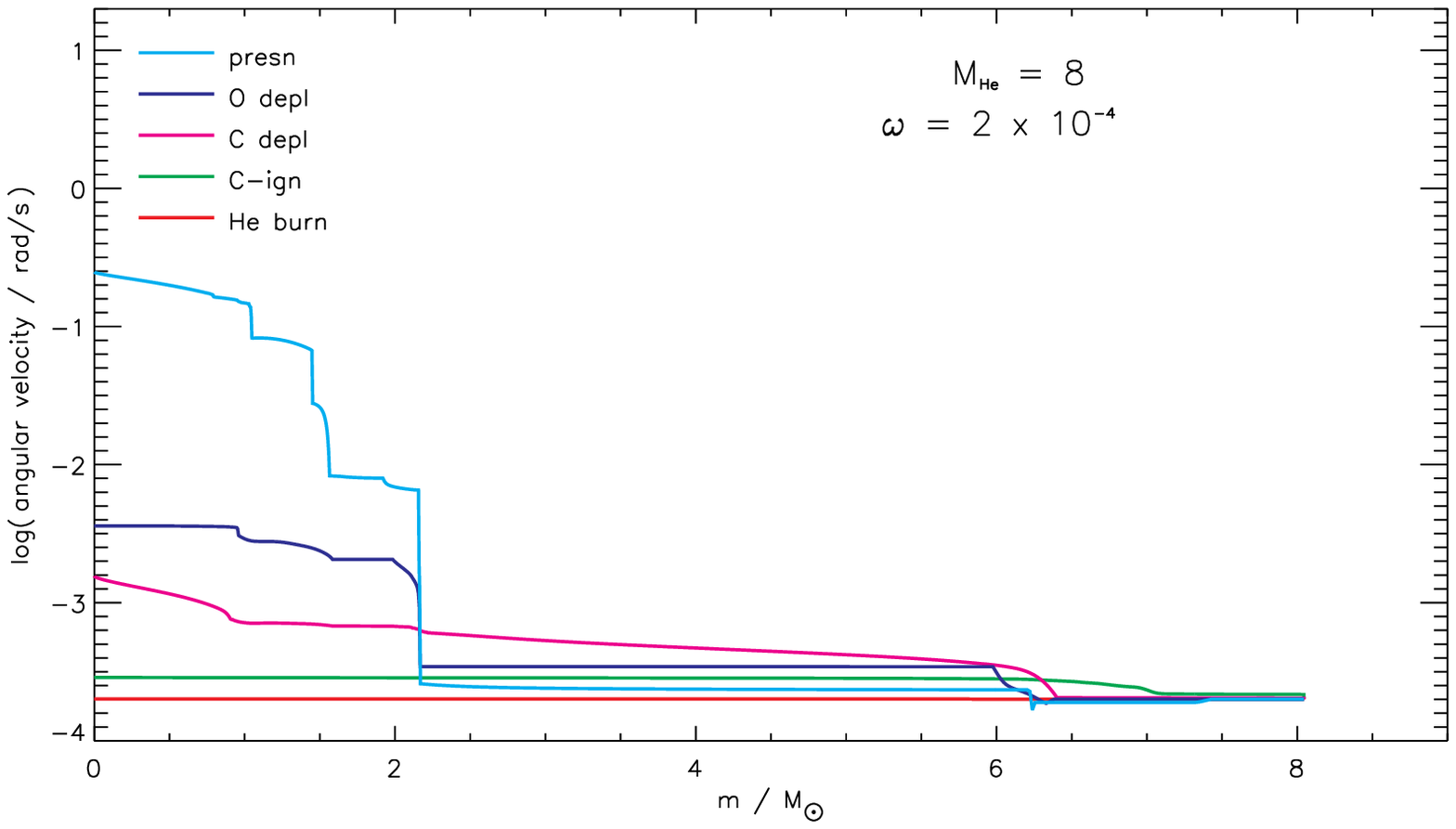}
\hfill
\includegraphics[width=0.475\textwidth]{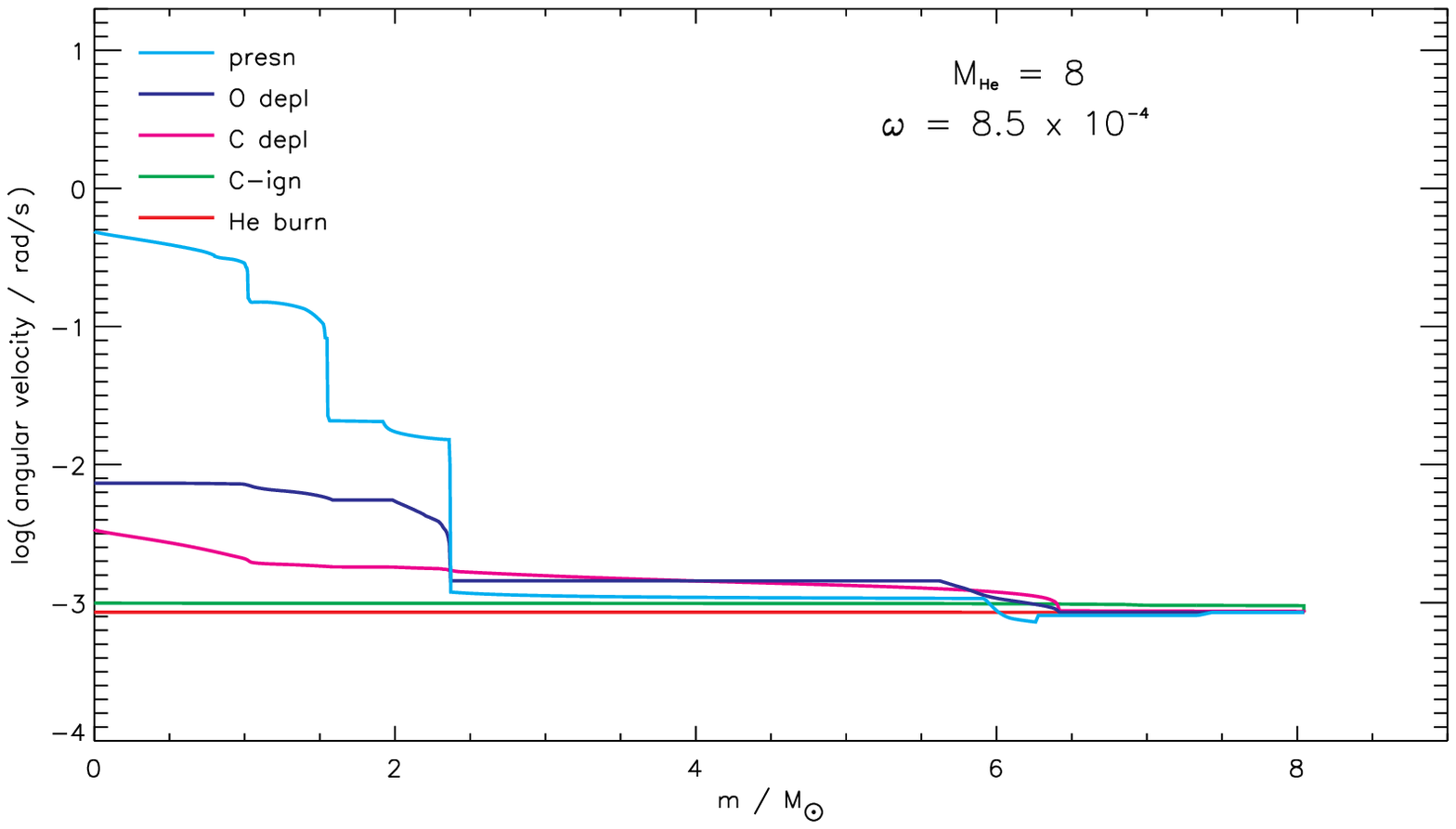}
\vskip 24pt
\includegraphics[width=0.475\textwidth]{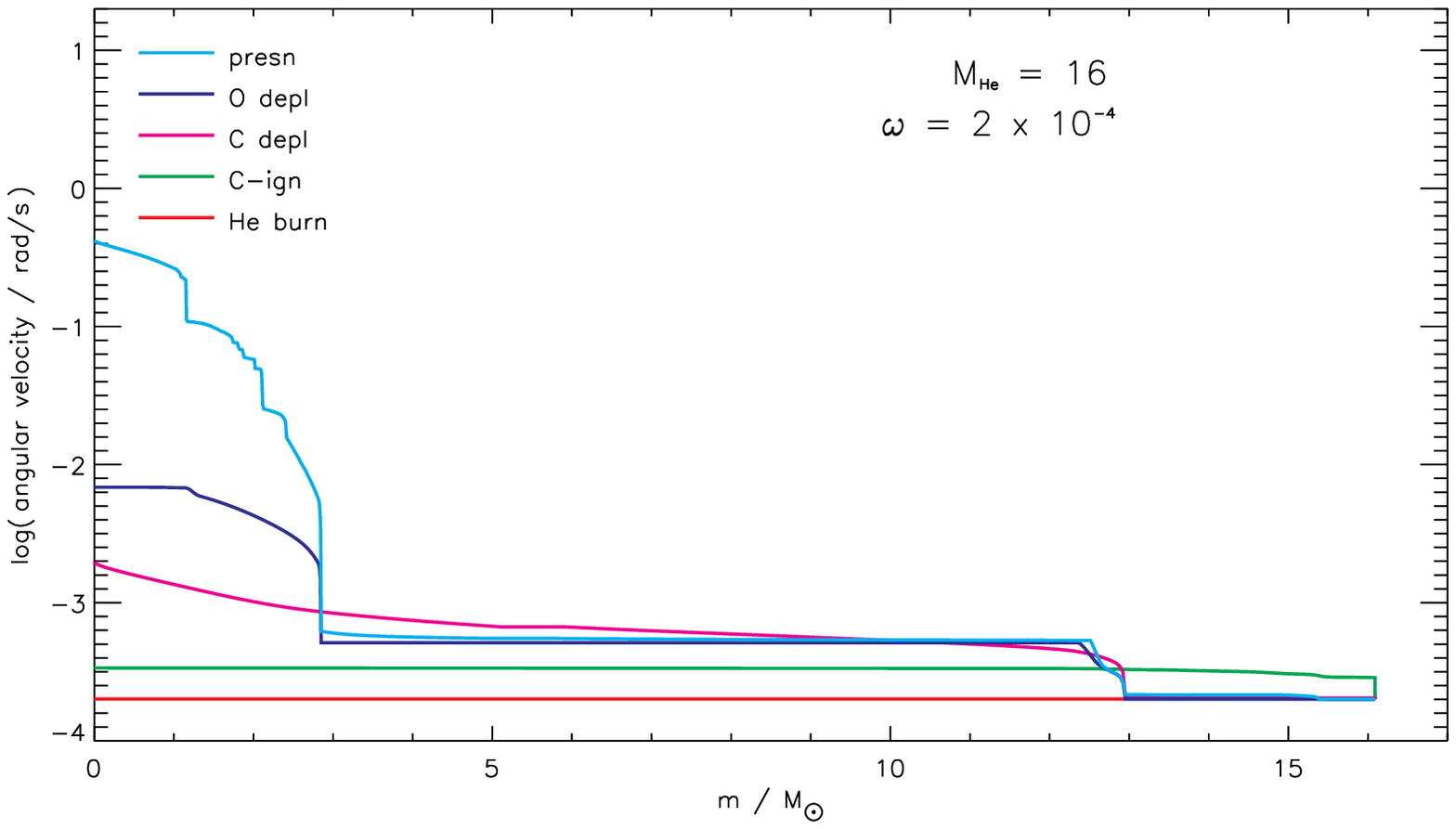}
\hfill
\includegraphics[width=0.475\textwidth]{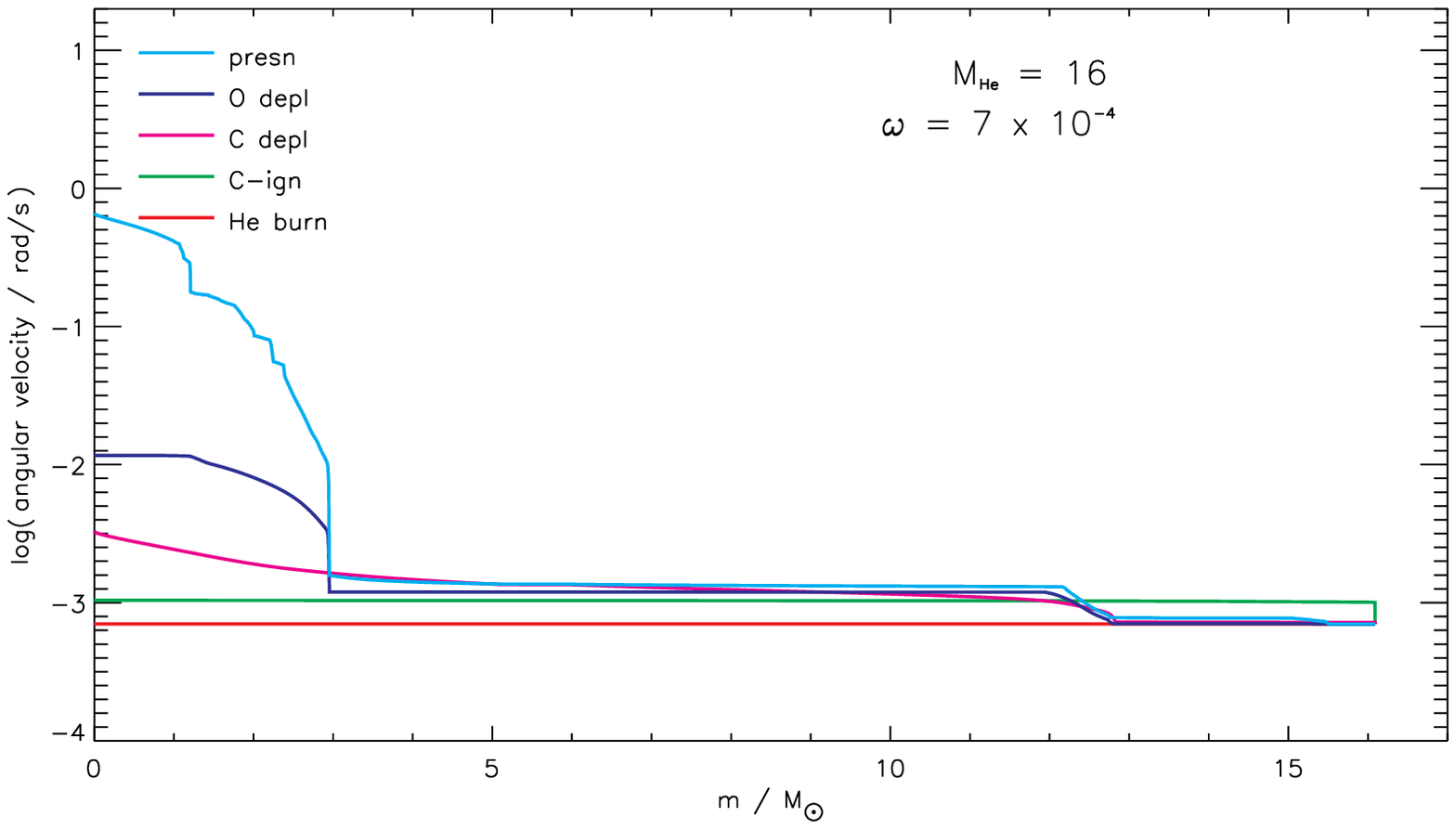}
\vskip 24pt
\caption{ History of the angular velocity for $8\,\Msun$ and
  $16\,\Msun$ helium cores in tidally locked binary systems. The two
  top panels show Models 8A (left) and 8B (right) and the bottom
  panels show Models 16A (left) and 16B (right). Note the tendency of
  all but the inner core to rotate with the (imposed) surface angular
  speed. At late times the core evolve rapidly and rotates
  differentially. \lFig{binaryw}}
\end{figure*}

\clearpage

\begin{figure*} 
\centering
\includegraphics[width=0.475\textwidth]{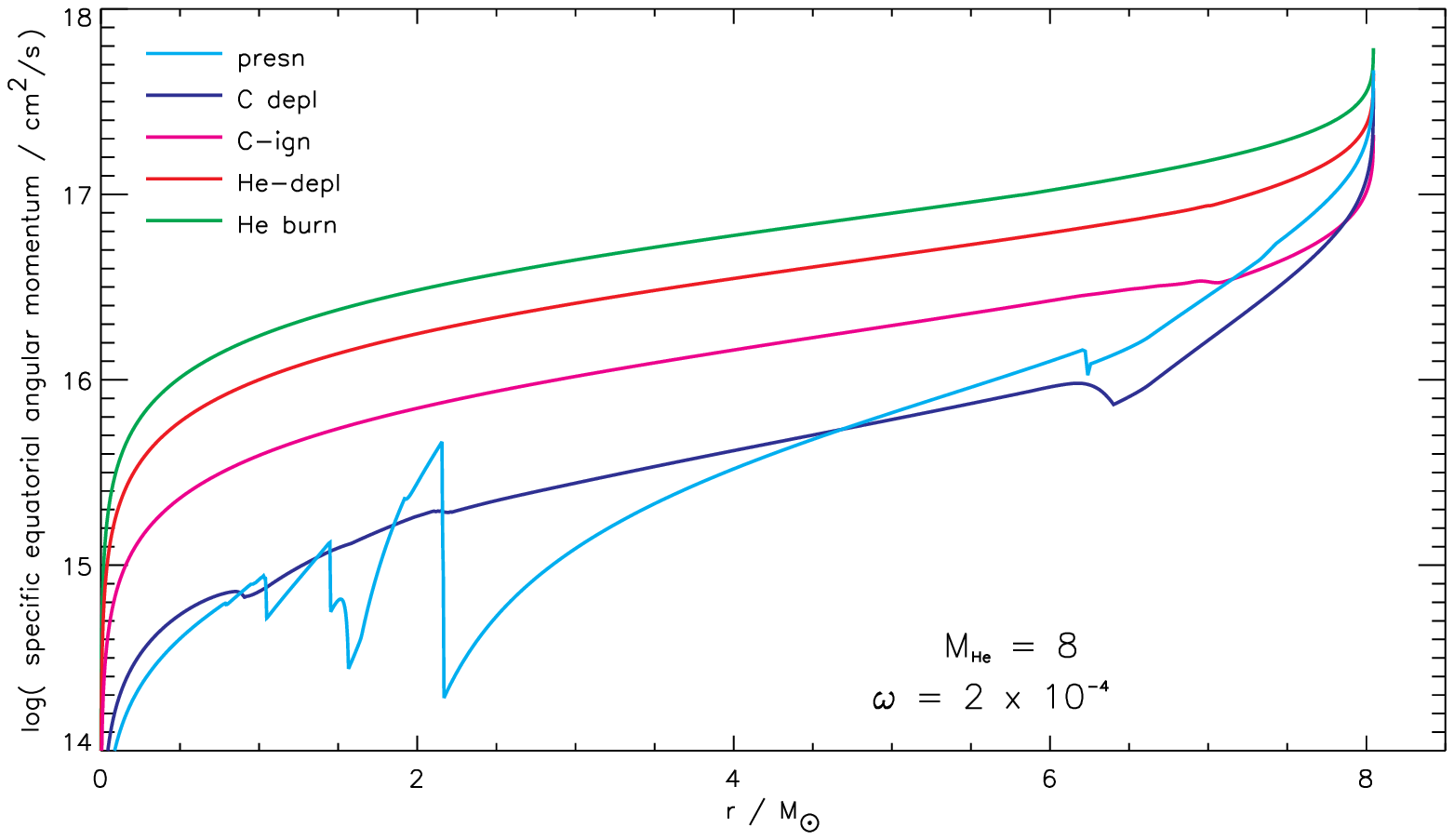}
\hfill
\includegraphics[width=0.475\textwidth]{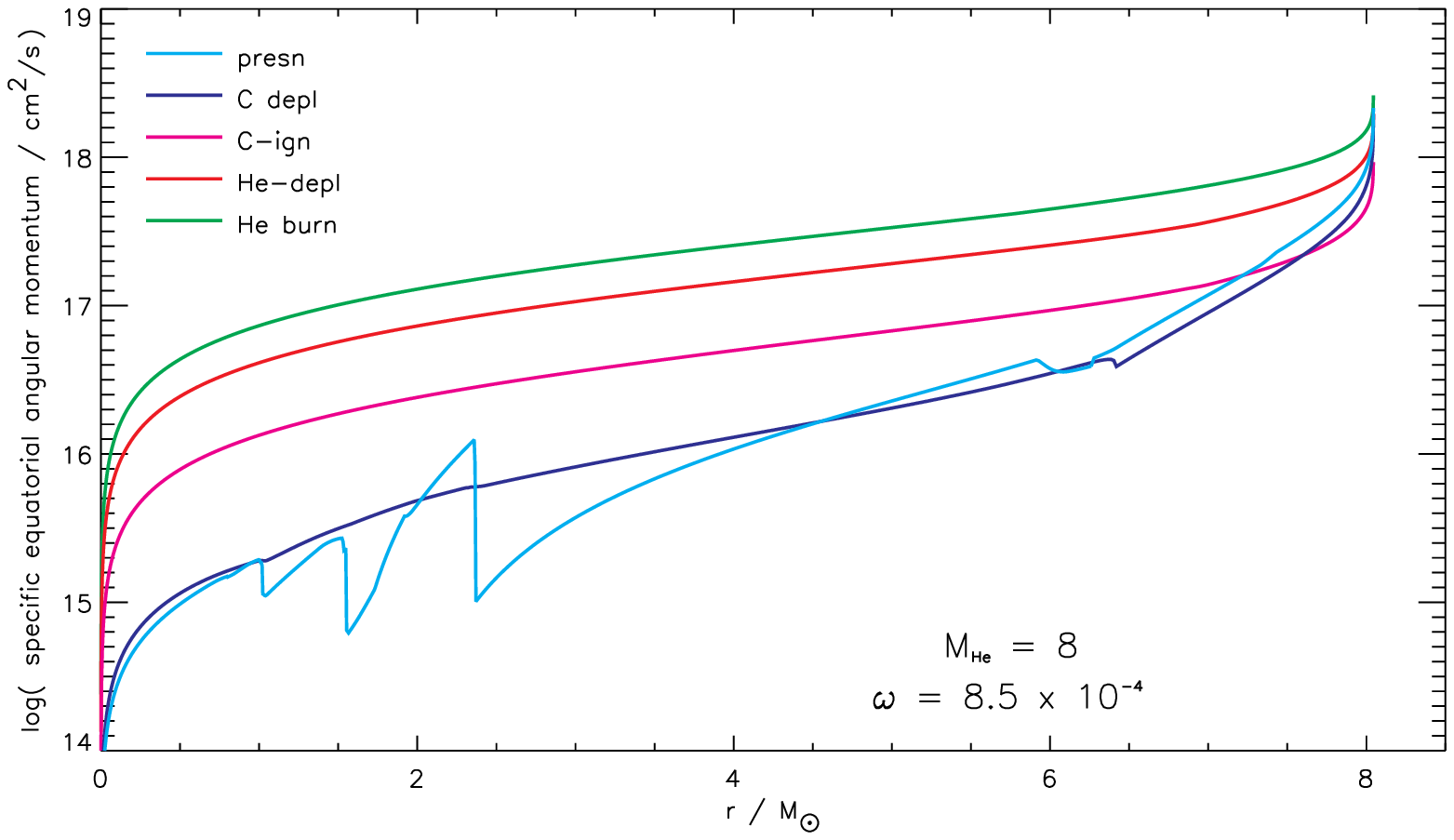}
\vskip 24pt
\includegraphics[width=0.475\textwidth]{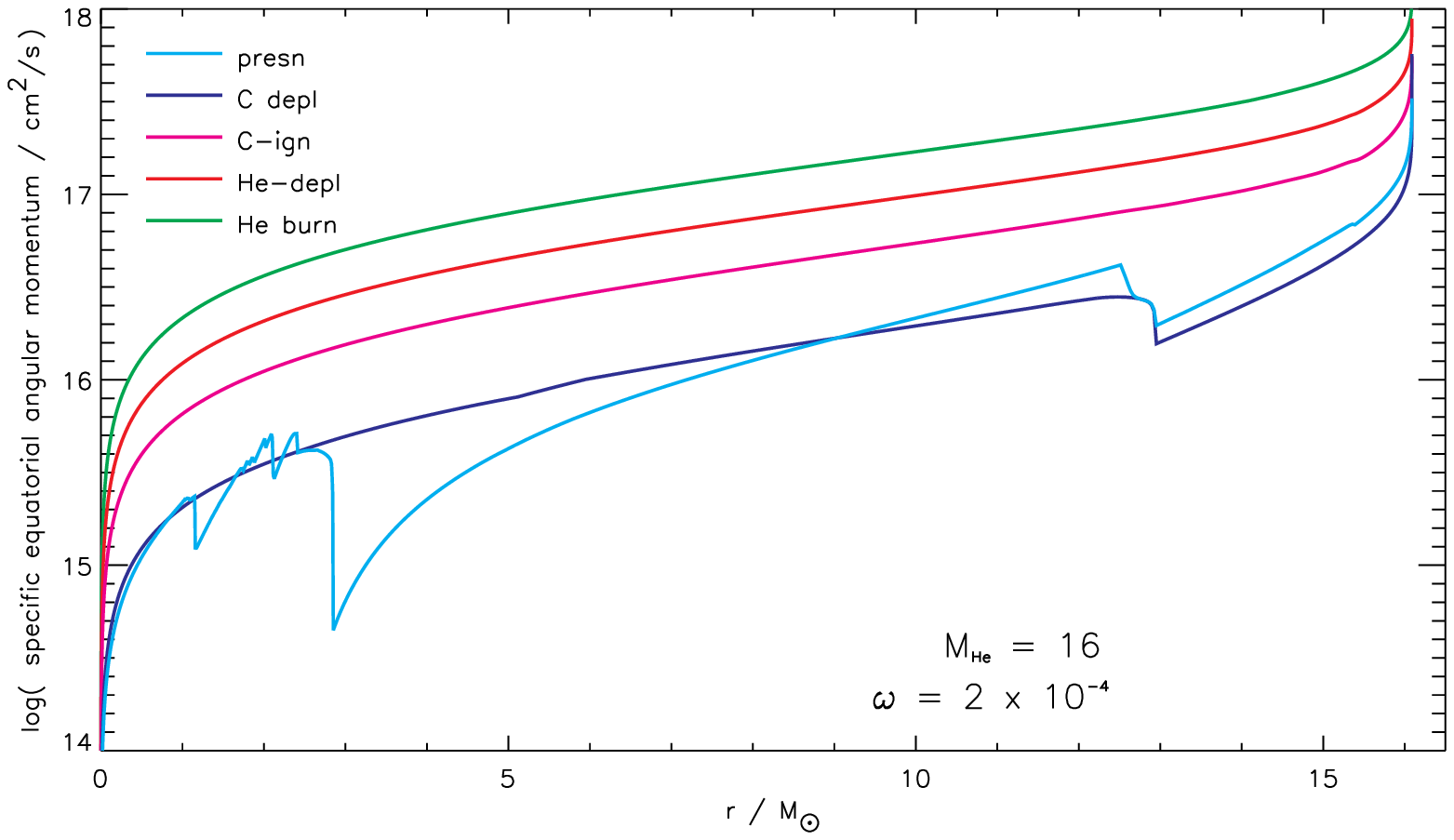}
\hfill
\includegraphics[width=0.475\textwidth]{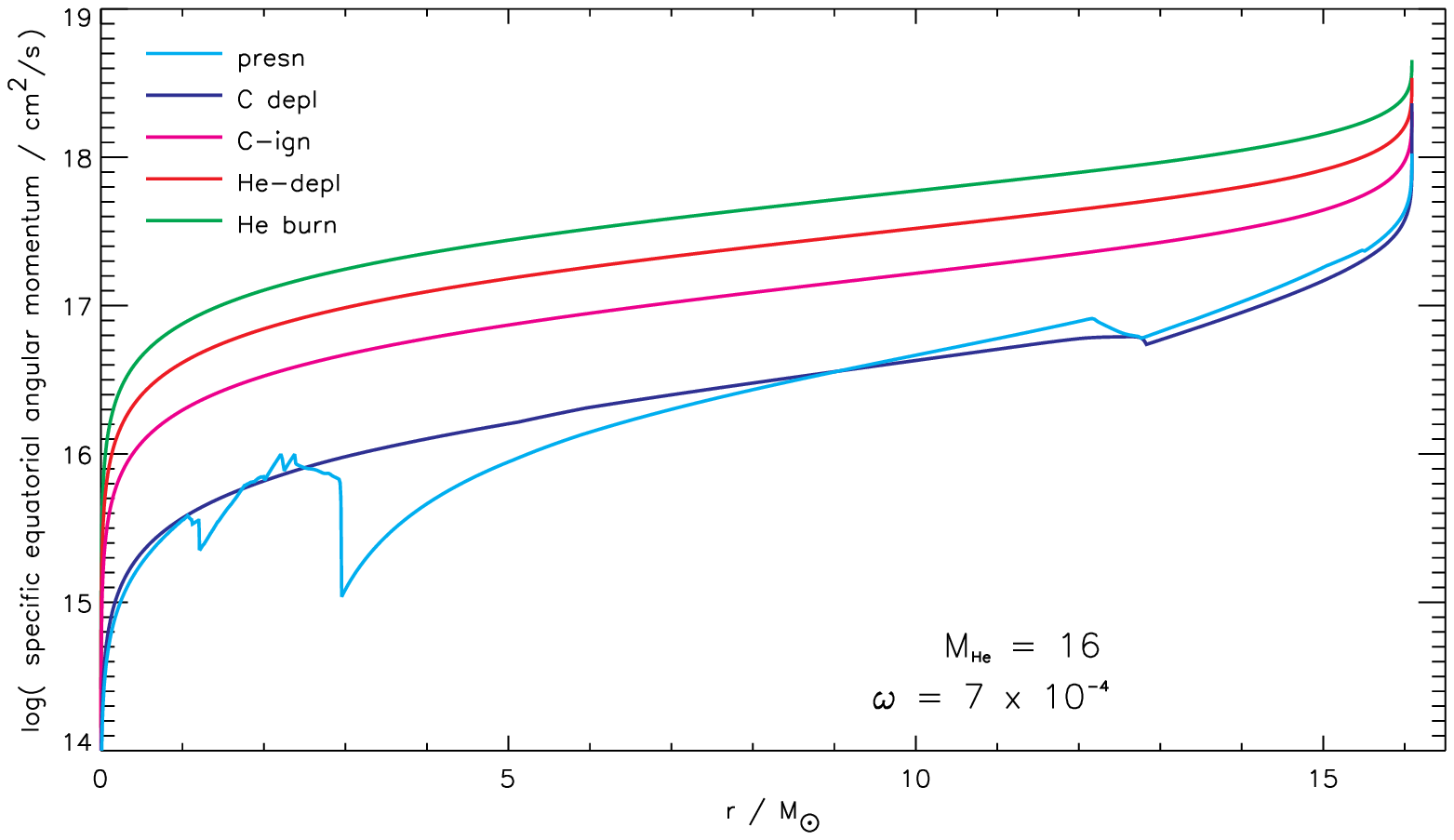}
\vskip 24pt
\caption{ History of the equatorial angular momentum for $8\,\Msun$
  and $16\,\Msun$ helium cores in tidally locked binary systems shown
  in \Fig{binaryw}.\lFig{binaryj}}
\end{figure*}

\clearpage

\begin{figure*} 
\centering
\includegraphics[width=0.475\textwidth]{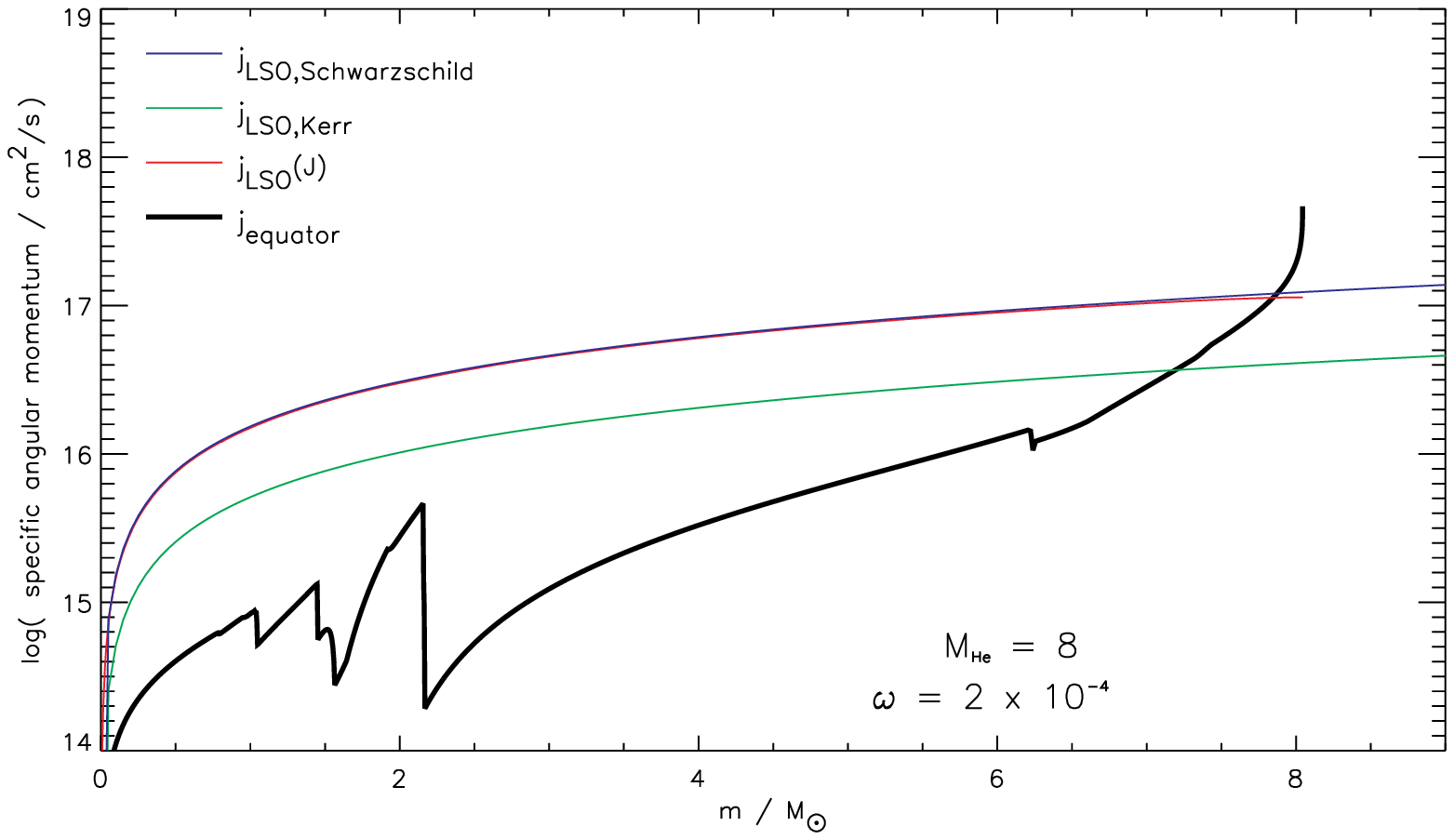}
\hfill
\includegraphics[width=0.475\textwidth]{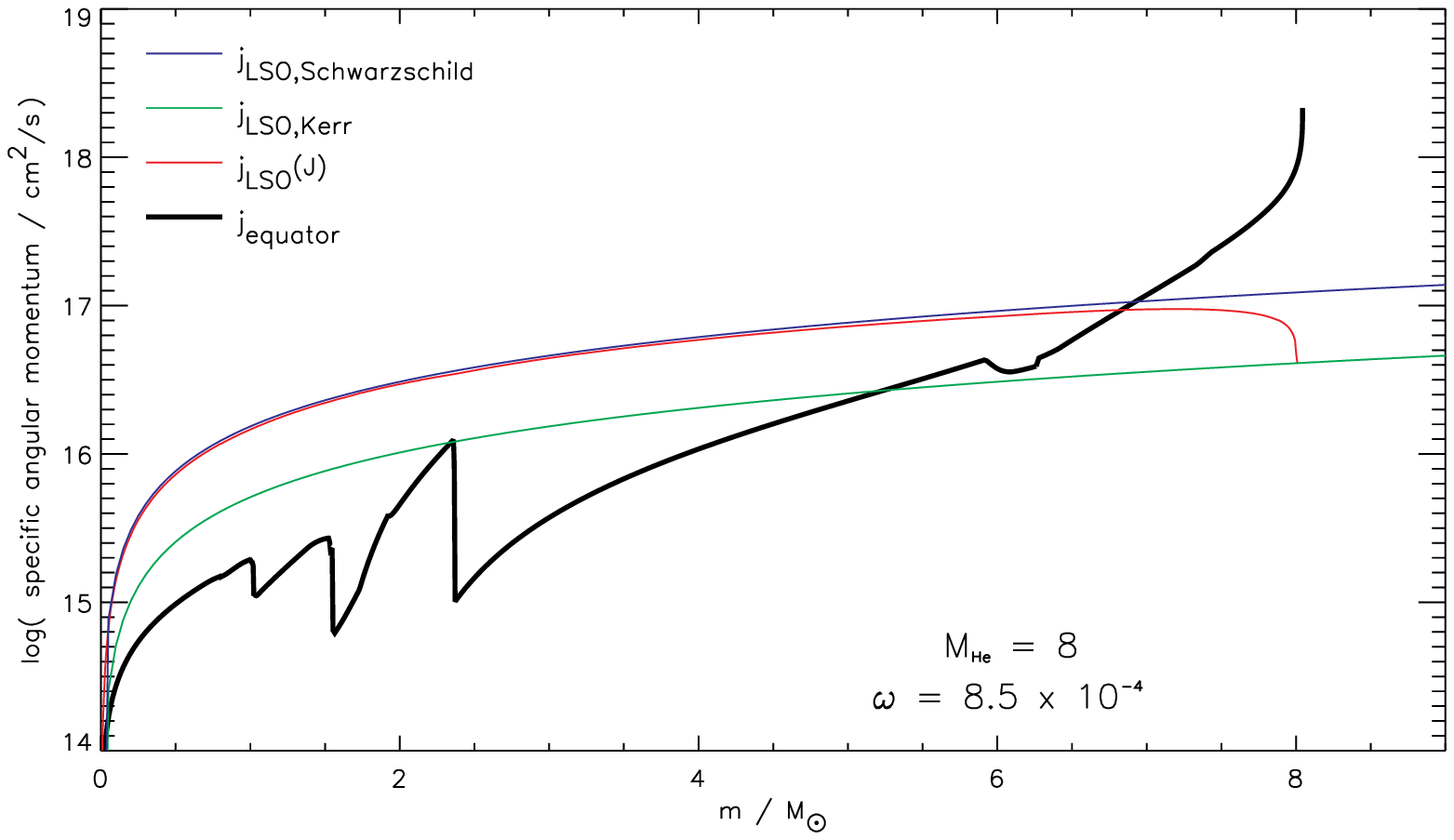}
\vskip 24pt
\includegraphics[width=0.475\textwidth]{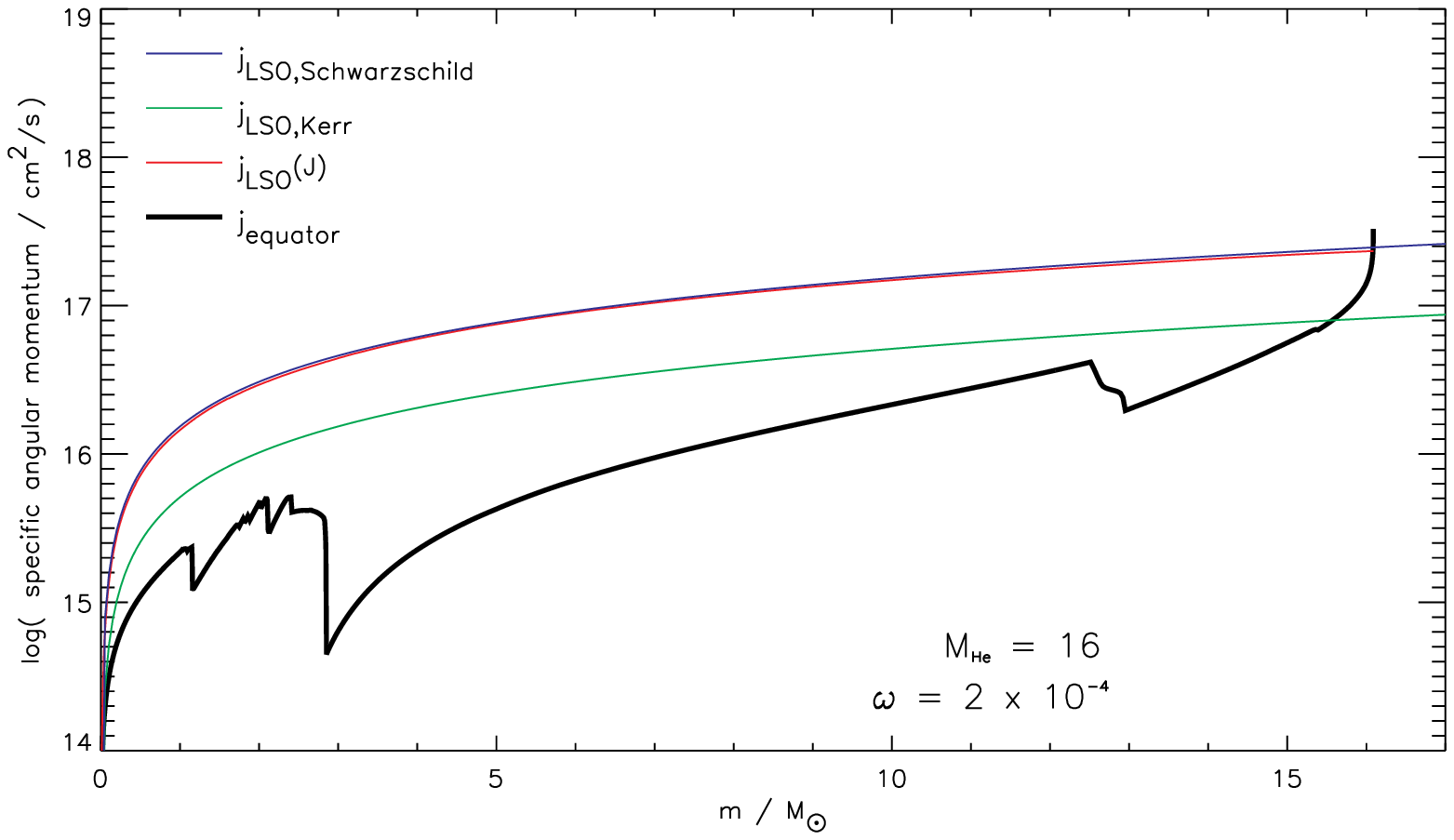}
\hfill
\includegraphics[width=0.475\textwidth]{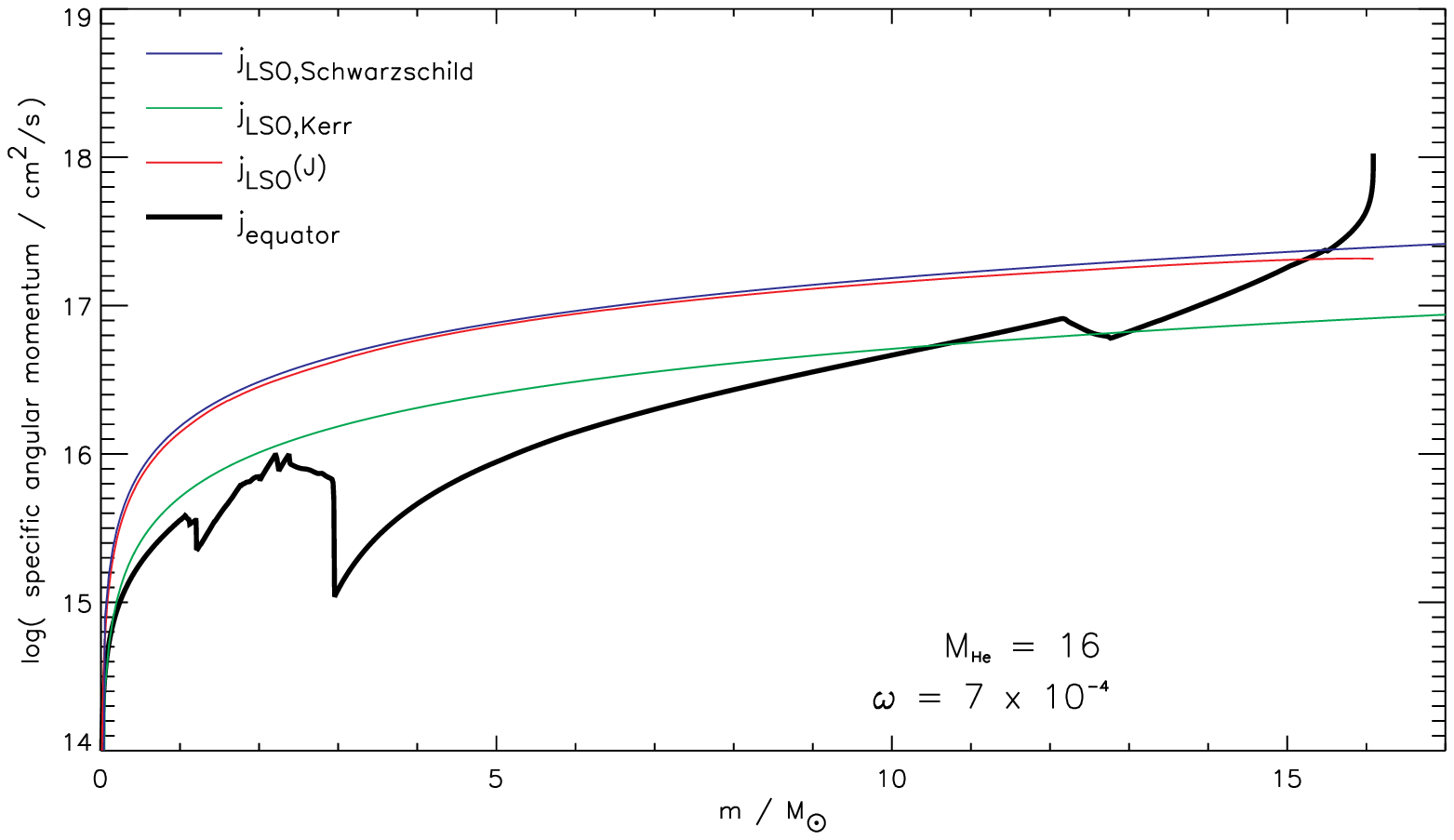}
\vskip 24pt
\caption{Final angular momentum for $8\,\Msun$ and $16\,\Msun$ helium
  cores in tidally locked binary systems compared to that required to
  form a disk around a black hole. Models 8B (upper right) clearly
  could make a disk and a rapidly rotating black hole. Model 16A
  (lower left) would not. Disk formation in Models 8A and 16B depend
  on the retention and collapse of a small bit of mass at the
  surface. \lFig{binaryjfin}}
\end{figure*}

\clearpage

\begin{deluxetable}{ccccccc}
\tablecaption{Summary of Model Characteristics}
\tablehead{
\colhead{Type} &
\colhead{Model} &
\colhead{$\Mdisk$} &
\colhead{$R$}  &
\colhead{$R/\vesc$} &
\colhead{$\Mdot$} &
\colhead{$1\,\%$ $\Mdot c^2$} 
\\
\colhead{}   &
\colhead{} &
\colhead{($\Msun$)}    &
\colhead{($\cm$)}  &
\colhead{($\Sec$)}    &
\colhead{($10^{-4}\,\Msuns$)} &
\colhead{($10^{48}\ergs$)}
}
\startdata
BSG     & V24   & 10.0  & 0.4 - 10 $\times 10^{12}$ & 20,000 &   5     & 9    \\
        & V36   & 10.2  & 1 - 50 $\times 10^{12}$   & 60,000 &   1     & 2    \\
RSG-loZ & O25   & 0.    & 1.2 $\times 10^{14}$      & 10$^7$  &  ...   & ...  \\
RSG-bin & S25   & 2.8   & 0.1 - 9 $\times 10^{13}$  & $3 \times 10^6$ & 0.01 & 0.02 \\
Pair SN & Z250A & 19.6  & 0.9 - 8 $\times 10^{13}$  & $3 \times 10^5$ & 0.7 & 1 \\
        & Z250B & 20.6  & 0.4 - 6 $\times 10^{13}$  & 10$^5$ &    2   &  4    \\
        & Z250C & 12.1  & 0.2 - 2 $\times 10^{13}$  & $3 \times 10^4$ & 4 &  8\\
WR-bin  & 8A    & 0.18  & 2.5 - 5 $\times 10^{10}$  & 100    &   20   &  40   \\
        & 8B    & 1.19  &  1 - 5 $\times 10^{10}$   & 100    &  100   & 200   \\
        & 16A   & 0.01  &  3 - 4 $\times 10^{10}$   & 100    &    1   &   2   \\
        & 16B   & 0.83  &  2 - 4 $\times 10^{10}$   &  40    &  200   & 300   \\
\enddata
\lTab{models}
\end{deluxetable}

\begin{deluxetable}{ccccccccc}
\tablecaption{Remnant Properties}
\tablehead{
\colhead{Type} &
\colhead{Model} &
\colhead{Period} &
\colhead{$\Mpresn$} &
\colhead{$M(j > \jcrit)$}  &
\colhead{$\BE(j > \jcrit)$} &
\colhead{$a$} &
\colhead{$\Mhecore$} &
\colhead{$\aHe$}
\\
\colhead{} &
\colhead{} &
\colhead{($\ms$)} &
\colhead{($\Msun$)}    &
\colhead{($\Msun$)}    &
\colhead{($10^{49}\,\erg$)}  &
\colhead{}    &
\colhead{($\Msun$)} &
\colhead{}  
} 
\startdata
BSG     & V24 & 5.4 &  23.96 & 14.0 - 24     &  2.1   & 1  & 8.32 & 0.13 \\
        & V36 & 3.7 &  35.85 & 25.7 - 35.9   &  1.2   & 1  & 14.97& 0.11 \\
RSG-loZ & O25 & 5.1 &  23.76 &   ...         &   ...  &0.52& 10.07& 0.12 \\
RSG-bin & R25 & 4.6 &  12.09 & 9.3 - 12.1    & 0.0025 & 1  & 9.19 & 0.13 \\
Pair SN & 250A& ... &   250  & 230.4 - 250   &  2.9   & 1  & 142  & 0.05 \\
        & 250B& ... &   250  & 229.4 - 250   &  8.2   & 1  & 166  & 0.09 \\ 
        & 250C& ... &   250  & 237.9 - 250   &  6.8   & 1  & 222  & 0.13 \\
WR-bin  & 8A  &  7.4&  8.043 & 7.86 - 8.04   &  1.0   &...& 8.04  & 0.25 \\
        & 8B  &  3.2&  8.043 & 6.85 - 8.04   & 7.8    &...& 8.04  & 1    \\
        & 16A &  2.5& 16.087 & 16.08 - 16.09 & 0.08   &...& 16.08 & 0.18 \\
        & 16B &  1.5& 16.087 & 15.26 - 16.09 & 12     &...& 16.08 & 0.48 \\
\enddata
\lTab{models1}
\end{deluxetable}

\end{document}